\documentclass[reprint,twocolumn,superscriptaddress,nofootinbib,amsmath,amssymb]{aastex63}
\usepackage{amsmath,amsfonts,amssymb}
\usepackage{colortbl}
\usepackage{times}

\newcommand{\ie}{i.e.,~}
\newcommand{\eg}{e.g.,~}

\begin{document}

\title{Bayesian analysis of neutron-star properties with parameterized
  equations of state: the role of the likelihood functions}

\author[0000-0002-9078-7825]{Jin-Liang Jiang}
\affiliation{Institut f\"ur Theoretische Physik, Goethe Universit\"at,
  Max-von-Laue-Str. 1, 60438 Frankfurt am Main, Germany}

\author[0000-0002-8669-4300]{Christian Ecker}
\affiliation{Institut f\"ur Theoretische Physik, Goethe Universit\"at,
  Max-von-Laue-Str. 1, 60438 Frankfurt am Main, Germany}

\author[0000-0002-1330-7103]{Luciano Rezzolla}
\affiliation{Institut f\"ur Theoretische Physik, Goethe Universit\"at,
  Max-von-Laue-Str. 1, 60438 Frankfurt am Main, Germany}
\affiliation{School of Mathematics, Trinity College, Dublin 2, Ireland}
\affiliation{Frankfurt Institute for Advanced Studies,
  Ruth-Moufang-Str. 1, 60438 Frankfurt am Main, Germany}

\begin{abstract}
We have investigated the systematic differences introduced when
performing a Bayesian-inference analysis of the equation of state of
neutron stars employing either variable- or constant-likelihood
functions. The former have the advantage that it retains the full
information on the distributions of the measurements, making an
exhaustive usage of the data. The latter, on the other hand, have the
advantage of a much simpler implementation and reduced computational
costs. In both approaches, the EOSs have identical priors and have been
built using the sound-speed parameterization method so as to satisfy the
constraints from X-ray and gravitational-waves observations, as well as
those from Chiral Effective Theory and perturbative QCD. In all cases,
the two approaches lead to very similar results and the $90\%$-confidence
levels are essentially overlapping. Some differences do appear, but in
regions where the probability density is extremely small and are mostly
due to the sharp cutoff set on the binary tidal deformability $\tilde
\Lambda \leq 720$ employed in the constant-likelihood analysis. Our
analysis has also produced two additional results. First, a clear inverse
correlation between the normalized central number density of a maximally
massive star, $n_{\rm c, TOV}/n_s$, and the radius of a maximally massive
star, $R_{\rm TOV}$. Second, and most importantly, it has confirmed the
relation between the chirp mass $\mathcal{M}_{\rm chirp}$ and the binary
tidal deformability $\tilde{\Lambda}$. The importance of this result is
that it relates a quantity that is measured very accurately,
$\mathcal{M}_{\rm chirp}$, with a quantity that contains important
information on the micro-physics, $\tilde{\Lambda}$. Hence, once
$\mathcal{M}_{\rm chirp}$ is measured in future detections, our relation
has the potential of setting tight constraints on $\tilde{\Lambda}$.
\end{abstract}

\section{Introduction}

Gravity compresses matter inside neutron stars (NSs) to densities several
times larger than the saturation number density of atomic nuclei $n_{\rm
  s}=0.16\,{\rm fm}^{-3}$. This makes NSs the densest known material
objects in the Universe. The existence of NSs has been conjectured
shortly after the experimental discovery of the neutron in the early 20th
century. Since then their existence has been confirmed by pulsar
observations and more recently by the direct detection of gravitational
waves from binary NS mergers by the LIGO/Virgo collaboration. Even after
several decades of extensive theoretical and observational efforts, our
knowledge about the interior composition and of such basic properties as
the mass-radius relation and the maximum mass of NSs is still
limited. The main reason for this is that first-principle calculations of
the properties of the equation of state (EOS) are currently not reliable
at the largest densities reached in NSs, although there are two limits --
either at very-low or very-high densities -- where our knowledge is on
firmer grounds~\citep[see, \eg][for an overview]{Dexheimer2019c}.

Indeed, at densities below and close to $n_{\rm s}$ Chiral Effective
Theory (CET) calculations~\citep{Hebeler:2013nza, Gandolfi2019,
  Keller:2020qhx, Drischler:2020yad} provide controlled predictions for
the EOS. In the opposite limit, \ie at densities beyond $\sim 40\,n_{\rm
  s}$, which are much larger than those reached even in the most massive
NSs, the EOS of Quantum Chromodynamics (QCD) becomes accessible via
perturbative methods~\citep{Freedman:1976ub,Vuorinen:2003fs,
  Gorda:2021kme, Gorda:2021znl}. However, at densities a few times larger
than $n_{\rm s}$, which are those realized inside NS cores, one either
needs to rely on model building~\citep[see, \eg][for some recent
  attempts]{Beloin2019, Bastian:2020unt, Traversi2020, Li2021,
  Demircik:2021zll, Ivanytskyi:2022wln} or on agnostic approaches that
are not based on microscopic models and that are similar in spirit what
done when modeling agnostic gravitational waveforms from binary
mergers~\citep[see, \eg][]{Bose2017}. What the two different 
approaches to model the EOS have in common is that both have to satisfy
the constraints provided by CET and perturbative QCD (pQCD), as well as
by the observational data of heavy pulsars and binary NS mergers.

Model-agnostic approaches to model the EOS have been employed extensively
in recent years and roughly fall into two categories, namely:
parameterized and non-parameterized methods. Non-parameterized approaches
include, for example, Gaussian processes to infer the
EOS~\citep{Landry2018, Gorda:2022jvk, Brandes:2022nxa, Legred2022},
machine learning~\citep{Morawski2020, Fujimoto2021, Soma2022b, Han2022a}
and non-parametric extensions of spectral expansion
method~\citep{Han2021, Han2022b}. On the other hand, there exist various
parameterized approaches that model the EOS by piecewise
polytropes~\citep{Read:2009a, Most2018, Zhao2018, OBoyle2020}, or via a
spectral-method representation of the EOS~\citep{Lindblom2010,
  Lindblom2018a}, as well as some hybrids of these~\citep{Jiang2020,
  Tang2021, Huth:2021bsp, Ferreira2021}.

In addition to the various EOS parameterization methods, there are
different ways to impose constraints from theory and astrophysical
measurements including their uncertainties that are typically provided in
form of confidence intervals. Arguably, the most popular way of imposing
constraints from observational data is the so-called \textit{Bayesian
  analysis} with \textit{variable likelihood} functions~\citep[see,
  \eg][for some recent works]{Greif2019, Jiang2020, Dietrich:2020efo,
  Tang2021, Raaijmaers2021b, Mohammad2021}, which builds a probabilistic
model that accounts for the measurement uncertainty in each of the
constraints imposed. A similar, but somewhat simpler approach, is to use
instead what we will refer to as \textit{sharp-cutoff method}~\citep[see,
  \eg][for some recent works]{Annala2019, Annala:2022, Altiparmak:2022,
  Ecker:2022, Ecker:2022dlg}, which makes no specific assumption about
how uncertainties are distributed, but imposes the constraints by simply
rejecting EOS models according to sharp-cutoff values that are, for
example, provided by the boundaries of some confidence interval. The
latter approach can be seen as equivalent to a Bayesian analysis in
which \textit{constant likelihood} functions for the constraints are used.

Both approaches have their advantages and disadvantages. The advantage of
a Bayesian analysis with variable likelihood functions is that -- when
such information is available -- it can fully account for uncertainties
in the measurements, even in the cases of bimodal distributions, which is
not possible in an analysis with constant likelihood functions. A
constant likelihood analysis, on the other hand, has the advantage of
being simpler to implement, more flexible in EOS modelling, and
numerically less expensive, while variable likelihood analysis can
quickly become numerically unfeasible, especially for models with many
parameters and many different constraints.

It is still unclear whether a full Bayesian analysis with a limited
parameter space using currently available observational data is able to
produce more accurate predictions than the simpler cutoff method. Hence,
the goal of this work is to investigate whether or not the two choices
for the likelihood can lead to significantly different results for the
EOS and hence for the NS properties. To enable a direct comparison of the
two approaches, we employ in both cases the sound-speed parameterization
method~\citep{Annala2019, Altiparmak:2022, Ecker:2022xxj} with identical
priors for the initial EOS ensemble. For simplicity, we restrict our
comparison to observational constraints only, meaning that the prior
ensembles in both cases are built by imposing the CET and perturbative
QCD constraints in a fixed-cutoff manner. Hence, the likelihood functions
of the CET and perturbative QCD constraints are constant, while that of
the observational constraints is suitably variable.

This paper is structured as follows. In Sec.~\ref{sec:method} we
introduce the two data-analysis approaches employed in this work. In
Sec.~\ref{sec:result} we present the results of this comparison and
identify various relations that are independent of the analysis scheme.
In Sec.~\ref{sec:conc_disc} we summarise and conclude.
Finally, in Appendix~\ref{app:HESS} we compare our results to a 
recent mass-radius measurement of a light compact object in HESS~J1731-347. 
Throughout the manuscript we use units in which the speed of light and
the Newton's gravitational constant are all set to unity, i.e., $c=1$ and 
$G=1$.

\section{Methods}
\label{sec:method}

\subsection{Equation of State Parametrization}

The EOS models we construct are a patchwork of several different
components, which we briefly summarize below and that follows a procedure
that is similar to the one discussed in Ref.~\citet{Altiparmak:2022}, to which
we refer the reader for details. At densities below $0.5\,n_s$, we use
the Baym-Pethick-Sutherland (BPS) prescription~\citep{Baym71b}. Such
prescription for the EOS we then extend in the range $n/n_s\in[0.5-1.1]$
with a single but random polytrope of the form
$p=Kn^{\Gamma}$~\citep{Rezzolla_book:2013}, where $\Gamma$ is uniformly
sampled in $[1.77, 3.23]$, such that the pressure resides entirely
between the soft and stiff model of Ref.~\citet{Hebeler2013}, while the
polytropic constant $K$ is fixed by matching to the BPS EOS at $0.5~n_s$.

At densities between $1.1\,n_s$ and $\approx 40\,n_s$ we use a continuous
array of piecewise-linear segments for the sound speed as function of the
chemical potential as starting point to construct thermodynamic
quantities. Throughout this work we use $N=11$ segments of the following
sound speed form
\begin{equation}
    \label{eq:cs2}
    c_s^2(\mu)=\frac{\left(\mu _{i+1}-\mu \right) c_{s,i}^2+\left(\mu -\mu
   _i\right) c_{s,i+1}^2{}}{\mu _{i+1}-\mu _i}\,,
\end{equation}
where the values of the chemical potential $\mu_0$ and the sound speed at
the first matching point, $c_{s,0}^2$, are determined by the
corresponding polytrope; the values of $\mu_{N}=2.6\,\rm GeV$, as well as
$c_{s,N}^2$ are determined by the perturbative QCD boundary conditions
discussed below. In our previous work~\citep{Altiparmak:2022} we randomly
sample the remaining values of $c_{s,i}^2$ and $\mu_i$. But in this work we
fix the chemical potentials that are not determined by boundary
conditions to be log-equidistantly separated at
\begin{equation}
  \label{eq:mui}
  \mu_i=1.02\left(\frac{2.6}{1.02}\right)^{{i}/{N}}\,{\rm
    GeV\,\quad for} \quad i=1,...,N-1\,,
\end{equation}
and only sample the corresponding $c_{s,i}^2$ uniformly in the range
$[0,1]$ set by thermodynamic stability and causality. We not that the
factor $1.02\,\rm GeV$ appearing in Eq.~\eqref{eq:mui} is determined by
the largest chemical potential reached by the polytrope at
$n=1.1\,n_s$. Using a relatively large number of $N=11$ segments with
log-equidistant chemical potentials allows us to ensure sufficient
resolution at densities relevant for the NS interior. Keeping the
matching chemical potentials at constant values allows us to reduce the
number of free parameters in the Bayesian analysis and make it
numerically feasible for the large number of segments we use.

Finally, at chemical potential $\mu=2.6\,\rm GeV$, corresponding to a
number density of $\approx 40\,n_s$, we impose the parameterized
next-to-next-to leading order (2NLO) perturbative QCD result
of~\citet{Fraga2014} for the pressure of cold quark matter
\begin{equation}\label{eq:pqcd_approx}
 p_{\rm QCD}(\mu, X)=\frac{3}{4\pi^2}\frac{\mu^4}{3^4}\left[c_1-\frac{d_1
     X^{-\nu_1}}{(\mu/\text{GeV})-d_2 X^{-\nu_2}}\right]\,,
\end{equation}
where $c_1=0.9008$, $d_1=0.5034$, $d_2=1.452$, $\nu_1=0.3553$,
$\nu_2=0.9101$, and the effective renormalization scale parameter $X$ is
chosen in the range $[1, 4]$.

There has been recent progress in extending Eq.~\eqref{eq:pqcd_approx} to
partial 3NLO~\citep{Gorda:2021znl, Gorda:2021kme} resulting in a small
correction to which our analysis is however not sensitive. Naively one
could expect that imposing the constraint \eqref{eq:pqcd_approx} at
$\lesssim 40\,n_s$ has little impact on the EOS at NS densities that are
$\lesssim 7\,n_s$. However, a number of recent studies have shown that
high-density QCD is indeed constraining, not only for the
EOS~\citep{Komoltsev:2021jzg, Gorda:2022, Somasundaram:2022}, but also
for related quantities, such as the sound speed and the conformal anomaly
inside massive NSs~\citep{Ecker:2022dlg}.

\subsection{Observational Constraints}

In addition to theoretical constraints, there exists a growing amount of
astrophysical data that further constrain the dense matter EOS and in
consequence the predictions for NS properties. Arguably the largest
impact have direct mass measurements from Shapiro delay of binary systems
that have massive radio pulsars like
PSR~J0740+6620~($M=2.08^{+0.07}_{-0.07}\,M_{\odot}$)
by~\citet{Fonseca2021},
PSR~J0348+0432~($M=2.01^{+0.04}_{-0.04}\,M_{\odot}$)
by~\citet{Antoniadis2013} and
PSR~J1614-2230~($M=1.908^{+0.016}_{-0.016}\,M_{\odot}$)
by~\citet{Arzoumanian2018}. These measurements set lower limits on the
maximum mass $M_{\rm TOV}$ of static NSs. We will therefore refer to
them collectively as \textit{TOV constraints} from here on.

In addition, the NICER experiment has provided combined mass and radius
measurements from the X-ray pulse-profile modelling of PSR~J0030+0451
by~\citet{MCMiller2019b}
($M=1.44^{+0.15}_{-0.14}~M_\odot,~R=13.02^{+1.24}_{-1.06}~{\rm km}$) and
by~\citet{Riley2019}
($M=1.34^{+0.15}_{-0.16}~M_\odot,~R=12.71^{+1.14}_{-1.19}~{\rm km}$), and
of PSR~J0740+6620 by~\citet{Miller2021}
($M=2.08^{+0.07}_{-0.07}~M_\odot,~R=13.7^{+2.6}_{-1.5}~{\rm km}$) and
by~\citet{Riley2021}
($M=2.072^{+0.067}_{-0.066}~M_\odot,~R=12.39^{+1.30}_{-0.98}~{\rm
  km}$). These measurements offer bounds for the minimum and maximum
radii of typical ($M\approx 1.4\,M_\odot$) and massive ($M\approx
2\,M_\odot$) NSs. We will refer to these collectively as \textit{NICER
  constraints}.

There also exist constraints on the tidal deformability of NSs in binary
systems that have been obtained from the direct gravitational-wave (GW)
detection by the LIGO/Virgo Collaboration from the merger event
GW170817~\citep{Abbott2017}. From this measurement, upper bounds on the
tidal deformability $\Lambda_{1.4}\leq 580$~\citep{Abbott2018b} of
individual NS with $M=1.4\,M_\odot$, as well as for the binary tidal
deformability parameter $\tilde{\Lambda}_{1.186}\leq
720$~\citep{Abbott2018a} and for the chirp mass
\begin{equation}
  \mathcal{M}_{\rm chirp}:=(M_1 M_2)^{3/5}(M_1 + M_2)^{-1/5} =
  1.186^{+0.001}_{-0.001}\,M_\odot \,,
\end{equation}
have been derived. We will then refer to these constraints obtained from
GW170817 as the \textit{GW constraints}.

Finally, there are other constraints that could be used but will not be
imposed here, either because they are theoretical constraints and hence
with model-dependent uncertainties~\citep[see, \eg][for some constraints
  on the minimum radii derived from prompt collapse of the binary to a
  black hole]{Bauswein2017b, Koeppel2019, Tootle2021}, or because the
measurement uncertainties are excessively large. In particular, there
exist predictions for the maximum mass by a number of groups on the basis
of the GW170817 event and the corresponding gamma-ray burst event
GRB170817A, namely $M_{\rm TOV}\lesssim
2.16^{+0.17}_{-0.15}\,M_{\odot}$~\citep{Margalit2017, Rezzolla2017,
  Ruiz2017,Shibata2019}. However, such upper bounds on $M_{\rm TOV}$ also
require a number of assumptions about the collapse to a black hole of the
merged object and about the ejected mass leading to the kilonova signal
of AT~2017gfo~\citep[see, \eg][]{Nathanail2021}.

We also note several recent NS-mass measurements have triggered
considerable interest. For example, observations of the black-widow
pulsar PSR~J0952-0607 has lead to an estimated mass of
$M=2.35\pm0.17~M_\odot$~\citep{Romani:2022jhd}, which exceeds any
previous measurements, including that of
PSR~J2215+5135~($M=2.27^{+0.17}_{-0.15}\,M_\odot$)
by~\citet{Linares2018}. Similarly, the recent discovery of a light
central compact object with a mass of $M=0.77^{+0.20}_{-0.17}~M_\odot$
within the supernova remnant HESS~J1731-347~\citep{Doroshenko2022} has
been interpreted as the lightest NS known or a ``strange star''; the
associated radius has been estimated to be $R=10.4^{+0.86}_{-0.78}~\,{\rm
  km}$. In Appendix~\ref{app:HESS} we will contrast this estimate with
the one that can be inferred from our Bayesian analysis.

In the following we will discuss how we implement the TOV, NICER, and GW
constraints in our Bayesian analysis, starting with the traditional way
that uses variable likelihood functions, followed by the simpler
sharp-cutoff method that uses constant step functions for the
likelihoods. We will refer to the former as the \textit{Variable
  Likelihood (VL)} analysis and to the latter as the \textit{Constant
  Likelihood (CL)} analysis. In each case, we employ more than $1.5
\times 10^5$ causal and thermodynamically consistent posterior EOSs that
satisfy the nuclear theory and perturbative QCD boundary conditions, as
well as the observational constraints from pulsars and GW measurements.

\subsection{Bayesian Analysis}

Bayesian analysis has been extensively used to infer the probability
distribution of unknown parameters by exploiting the knowledge of some
measured variables. This is accomplished by constructing likelihood
functions that correlate unknown parameters and measured variables. In
this approach, the probability of a certain set of parameters is given by
the posterior probability
\begin{equation}
  \label{eq:total_likelihood}
 P(\vec{\theta}|\boldsymbol{d}) = \frac{L(\boldsymbol{d}|\vec{\theta})\,
   \pi(\vec{\theta})}{Z}\,,
\end{equation}
where the vector $\vec{\theta}$ collects all the free parameters,
$\boldsymbol{d}$ denotes the imposed measurement data,
$\pi(\vec{\theta})$ is the so-called ``prior'', and the normalization
constant $Z$ is also called the ``evidence''. In our case, the likelihood
$L$ can be written as the product of the likelihoods of the three
constraints we impose
\begin{equation}
  \label{eq:likelihood}
  L(\boldsymbol{d}|\vec{\theta})=\mathcal{L}^{\rm TOV}(\vec{\theta}_{\rm
    EOS})\, \mathcal{L}^{\rm NICER}(\vec{\theta}) \, \mathcal{L}^{\rm
    GW}(\vec{\theta})\,,
\end{equation}
where the individual likelihood functions $\mathcal{L}^{\rm TOV},
\mathcal{L}^{\rm NICER}$, and $\mathcal{L}^{\rm GW}$ will be defined in
detail below and the free-parameters vectors are given by
\begin{equation}
\vec{\theta}:=\vec{\theta}_{\rm EOS} \cup \{\mu^{\ell}_{c}| \,\ell=
1,2,3,4\}\,,
\end{equation}
and
\begin{equation}
\vec{\theta}_{\rm EOS}:= \{c^2_{{\rm s},i}| \,i= 1,2,...,N-1\} \cup
\Gamma\,.
\end{equation}
The four chemical potentials $\mu^{\ell}_{c}$ refer to the values at the
center of the NSs, two of which refer to pulsars with simultaneous
mass-radius measurements, while the other two to the two NSs composing
the GW170817 event. The three types of parameters in this construction,
\ie $c^2_{{\rm s}}, \Gamma$, and $\mu^{\ell}_{c}$, are uniformly
distributed in the following ranges
\begin{equation}
  \label{eq:prior}
  c^2_{{\rm s}, i}\in [0, 1],\quad \Gamma \in [1.77, 3.23],\quad
  \mu^{\ell}_{c} \in [1, 2.1]~\rm GeV\,.
\end{equation}
For the parameter sampling we use the
\texttt{Pymultinest}~\citep{Buchner2016} algorithm implemented in
\texttt{Bilby}~\citep{Ashton2019}.

\subsection{Variable-Likelihoods Analysis}

For simplicity, to approximate the mass measurements of PSR~J0740+6620,
PSR~J0348+0432, and PSR~J1614-2230, we use Gaussian distributions and the
Cumulative Density Function (CDF) of each Gaussian is then used to build
the likelihood contribution of the corresponding pulsar, namely
\begin{equation}
    \label{eq:like_pulsar_i}
	\mathcal{L}^{\rm TOV}_{\rm i}(\vec{\theta}_{\rm EOS}) =
        \frac{1}{2}\left[1+\text{erf}\left(\frac{M_{\rm
              TOV}(\vec{\theta}_{\rm
              EOS})-\bar{M}_i}{\sqrt{2}\sigma_i}\right)\right]\,,
\end{equation}
where
\begin{equation}
  \text{erf}(x):=\frac{2}{\sqrt{\pi}}\int_0^x e^{-t^2}dt\,,
\end{equation}
is the error function, while $\bar{M}_i$ and $\sigma_i$ ($i=1,2,3$) are
the mean and the standard deviation of the $i$-th mass measurement,
respectively. The total likelihood for the \textit{TOV constraints} is
then given by the product of the individual likelihoods of these three
pulsars 
\begin{equation}
    \label{eq:like_pulsar_total}
    \mathcal{L}^{\rm TOV}(\vec{\theta}_{\rm EOS}) := \prod^{3}_{i=1}
    \mathcal{L}^{\rm TOV}_{\rm i}(\vec{\theta}_{\rm EOS}).
\end{equation}

The probability distribution function of mass and radius of PSR
J0030+0451\footnote{We have employed the best-fit
\href{https://zenodo.org/record/5506838}{ST+PST}.} and PSR
J0740+6620\footnote{We have employed the data file
\href{https://zenodo.org/record/4697625\#.YKMcuy0tZQJ}{STU/NICERxXMM/FI\_H/run10}.}
are all estimated by the Kernel Density Estimation (KDE)
\citep{Scott1992} with a Gaussian kernel using the released posterior
$(M,R)$ samples $\boldsymbol{S}_k$. The likelihood of \textit{NICER
  constraints} can then be represented as the product of the KDE of each
pulsar, namely 
\begin{equation}
    \label{eq:like_psr}
    \mathcal{L}^{\rm NICER}(\vec{\theta}) := \prod^{2}_{k=1}
    \text{KDE}_k\left(M(\vec{\theta}_{\rm EOS}, \mu^{k}_{c}),
    R(\vec{\theta}_{\rm EOS}, \mu^{k}_{c})| \boldsymbol{S}_k\right)\,,
\end{equation}
where $ \vec{\theta}_{\rm EOS} \cup \{\mu^{k}_{c}|\, k=1,2\}$ are parameters
needed to construct this likelihood and they constitute subset of $\vec{\theta}$
, while the mass $M$ and the radius $R$ of the $k$-th pulsar are
functions of the sampled EOS parameters $\vec{\theta}_{\rm EOS}$ and of
the central chemical potential parameters $\mu^{k}_{c}$.

When imposing a measurement of a single GW event, the corresponding 
likelihood can be expressed as~\citep{Abbott2020d}
\begin{multline}
    \label{eq:like_gw}
    \mathcal{L}^{\rm GW}(\boldsymbol{d}|\vec{\theta}_{\rm GW}, \boldsymbol{W})
    \propto \\ \prod_{i}^{N_d} \exp{ \left[-2\int_0^{\infty}
        \frac{|\tilde{d}_i(f)-\tilde{h}_i(f; \vec{\theta}_{\rm GW},
          \boldsymbol{W})|^2}{S^i_n(f)}\, df\right]}\,,
\end{multline}
where $\boldsymbol{W}$ represents the waveform model, $i$ and $N_d$ 
denote the $i$-th and total number of GW detectors that have detected 
this event respectively, while $S^i_n$, $\tilde{d}_i$ and $\tilde{h}_i$ 
represent the power spectral density of the detector noise, the detected 
strain signal, and the expected strain from the waveform model,
respectively. 

The vector $\vec{\theta}_{\rm GW}$ includes parameters that are
particularly useful to infer EOS properties, namely
\begin{equation}
\vec{\theta}^{\rm EOS}_{\rm GW} := \{M^{1}_{d}, M^{2}_{d}, \Lambda_1,
\Lambda_2 \}\,,
\end{equation}
where $M^{j}_{d}$ ($j=1,2$) are the individual masses in the detector
frame, while the tidal deformabilities of the two NSs are computed as
\begin{equation}
    \label{eq:def_lambda}
       \Lambda_j :=\frac{2}{3}k_2\left(\frac{R_j}{M^{j}_{s}}\right)^5
       =\Lambda_j(\mu^{j}_{c}, \vec{\theta}_{\rm EOS})\,.
\end{equation}
In the expression above, $R_j$ is the radius of the $j$-th NS in the
binary, $k_2$ is the second tidal Love number, $M^{j}_{s}$ is the mass
in the source frame, and $\mu^{j}_{c}$ denotes the central chemical
potential. The relation between the masses in the two frames is simply
mediated by the cosmological redshift $z$, namely
\begin{align}
  \label{eq:cal_ml}
	M^{j}_{d}&=(1+z)M^{j}_{s}(\mu^{j}_{c}, \vec{\theta}_{\rm
          EOS})\,,
\end{align}
where $z=0.0099$ for the GW170817 event~\citep{Abbott2018a}.
To this scope, we use the interpolated likelihood function for GW170817
that is implemented in \texttt{Toast}~\citep{Vivanco2020} and that is
marginalized over all the ``nuisance parameters''
$\vec{\theta}^{\rm nui}_{\rm GW} := \vec{\theta}_{\rm
  GW}\setminus\vec{\theta}^{\rm EOS}_{\rm GW}$ in Eq.~(\ref{eq:like_gw})
to construct the likelihood for the \textit{GW constraint}. These
nuisance parameters consist of a dozen of additional parameters that are
useful when performing a Bayesian analysis on GW-emitting binaries but
that are not relevant for our EOS inference. Including these parameters
would significantly slow down the sampling process and hence we use the
likelihood of \texttt{Toast} that marginalizes over these parameters.

\subsection{Constant-Likelihoods Analysis}

In the CL analysis we use the same set of astrophysical data as in the VL
analysis, but the corresponding likelihood functions are constructed
differently. More specifically, the likelihood functions are implemented
as Heaviside step-functions of the form 
\begin{align}
  \label{eq:H}
  H_{\pm} &= H_{\pm} \left(x_i(\vec{\theta}_{\rm EOS})-x_{i}^{\rm
    cut}\right) \nonumber \\
  &=
  \begin{cases}
    1,&  \qquad {\rm for}~\pm \left[ x_i(\vec{\theta}_{\rm EOS})-x_{i}^{\rm cut}\right]\geq 0\,, \\
    0,&  \qquad {\rm else}\,,
  \end{cases}
\end{align}
where $x_i(\vec{\theta}_{\rm EOS})$ is a quantity constructed from the
prior of EOS parameters $\vec{\theta}_{\rm EOS}$ and is either
constrained from above ($-$) or from below ($+$) by some sharp-cutoff
value $x_i^{\rm cut}$ obtained from the $i$-th measurement of this
quantity. In the following, in order to construct the CL functions, we
will use the same sharp-cutoff values $x_i^{\rm cut}$ for the
astrophysical constraints employed by~\citet{Altiparmak:2022}, which we
will discuss in detail below. On the other hand, for the EOS
parameterization, the priors of the EOS parameters and the theoretical
constraints from pQCD and CET will be the same as those in the VL
analysis discussed in the previous section.

For the \textit{TOV constraints}, we choose a single lower bound of
$M_{\rm TOV}\geq2.01\,M_{\odot}$, that is motivated by the mass
measurements of PSR J0740+6620 and PSR J0348+0432. The likelihood
function that implements this constraint is then simply given by
\begin{equation}
	\mathcal{L}^{\rm TOV}(\vec{\theta}_{\rm EOS})=H_{+}\left(M_{\rm
          TOV}(\vec{\theta}_{\rm EOS})-2.01\,M_\odot \right)\,.
\end{equation}

For the \textit{NICER constraints}, most relevant are the lower bounds on
the NS radii since the upper bounds are typically set by the
\emph{binary} tidal deformability for the GW170817 event
\begin{equation}
  \label{eq:Lambda}
  \tilde{\Lambda}:=\frac{16}{13}
  \frac{\left(12M_2+M_1\right)M_{1}^4\Lambda_1+\left(12M_1+M_2\right)M_{2}^4\Lambda_2}
       {\left(M_1+M_2\right)^5}\,.
\end{equation}
as these are effectively more constraining. It is also important to note
that measurement of NS masses $\gtrsim 2.2\,M_\odot$ make the
minimum-radius bounds obtained by NICER essentially ineffective, as
recently shown in~\citet{Ecker:2022dlg}. In practice, we choose a lower
bound of $R>10.8\, \rm km$ at $M=1.1\,M_{\odot}$ and a lower bound of
$R>10.75\, \rm km$ at $M=2.0\, M_{\odot}$ to approximate the radius
measurements of PSR J0030+0451 and PSR J0740+6620, respectively. The
corresponding likelihood function can then be written as
\begin{multline}
  \mathcal{L}^{\rm NICER}(\vec{\theta}_{\rm
    EOS})=H_{+}\left(R_{2.0}(\vec{\theta}_{\rm EOS})-10.75\,{\rm
    km}\right)\\ \times H_{+}\left(R_{1.1}(\vec{\theta}_{\rm
    EOS})-10.8\,{\rm km}\right)\,.
\end{multline}

Finally, as \textit{GW constraint}, we impose the upper bound $\tilde
\Lambda \leq 720$ derived from GW170817 only for a low-spin
prior.\footnote{Large spins are of course possible in a binary
merger~\citep[see][for a discussion of the possible ranges in the spin
  and mass ratios]{Most2020c}, but their inclusion would require a
consistent model of our stellar equilibria, which are here treated as
non-rotating.} The likelihood can be expressed as
\begin{equation}
	\mathcal{L}^{\rm GW}(\vec{\theta}_{\rm
          EOS})=H_{-}\left(\tilde{\Lambda}_{1.186}(\vec{\theta}_{\rm
          EOS})-720\right)\,,
\end{equation}
which has to be evaluated at a fixed chirp mass $\mathcal{M}_{\rm
  chirp}=1.186\,M_\odot$ and mass ratios $q:={M_1}/{M_2}\in [0.7, 1]$, as
imposed by the GW170817 detection.

\begin{figure*}[htb]
    \center
    \includegraphics[width=0.48\textwidth]{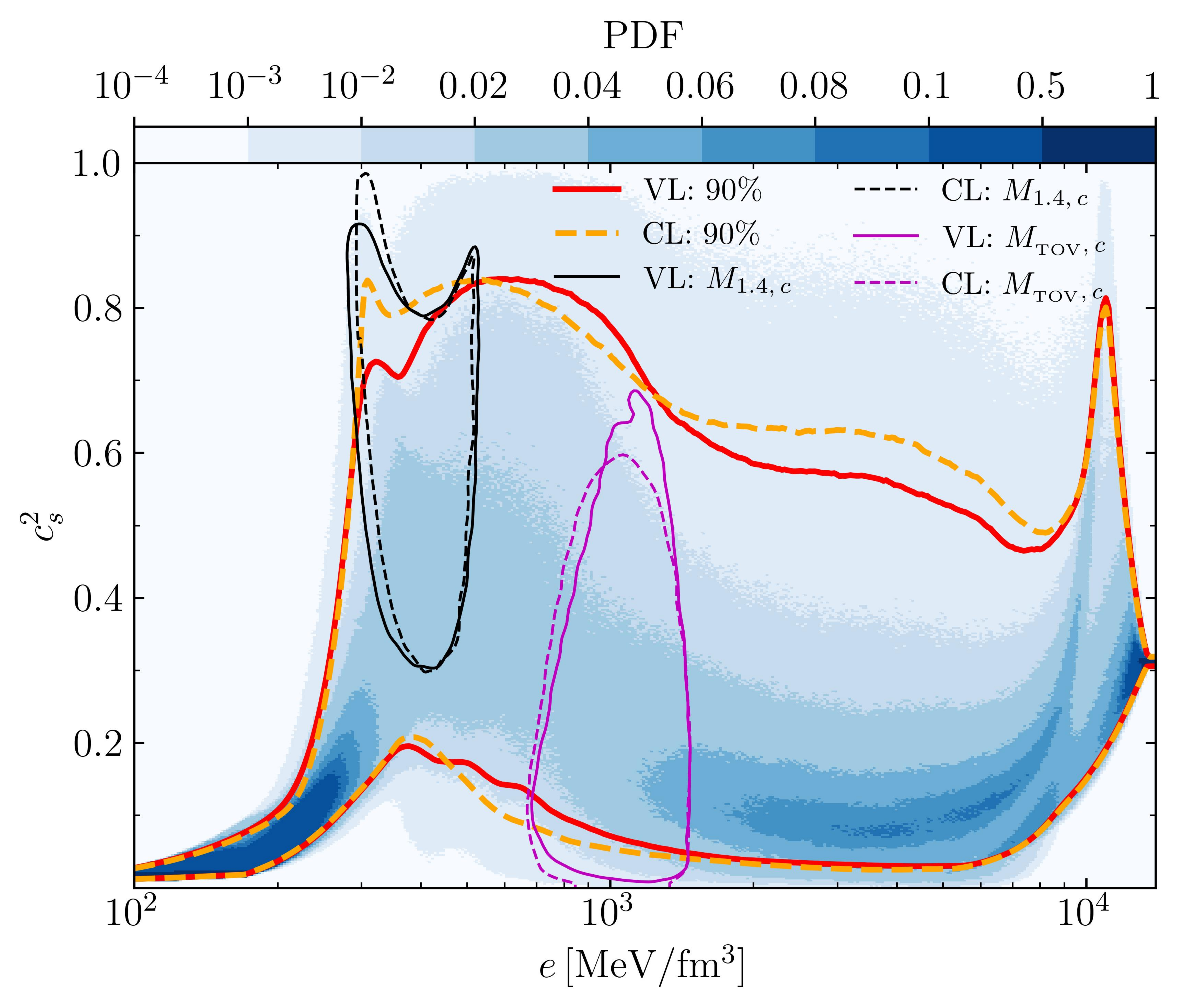}
    \includegraphics[width=0.48\textwidth]{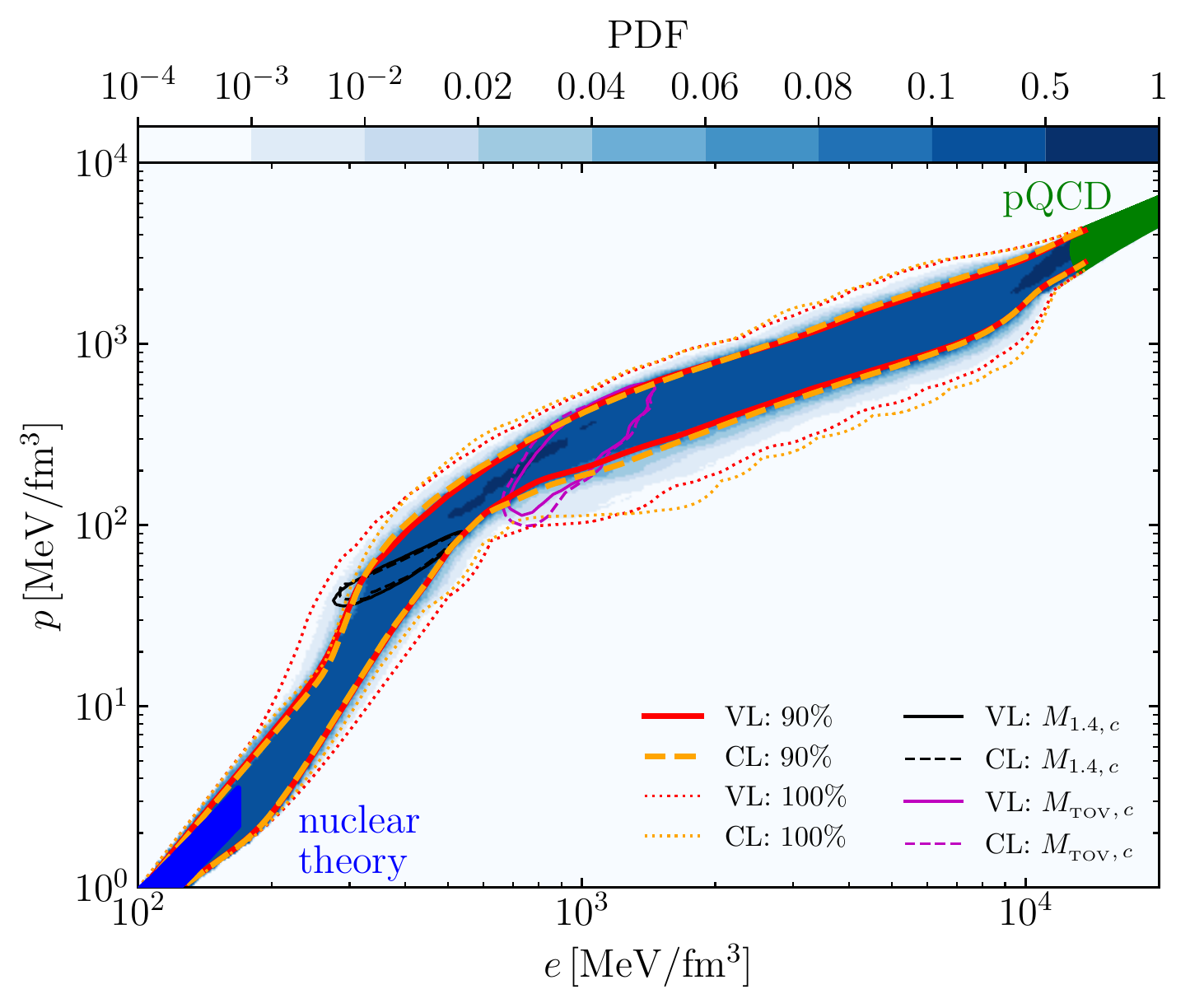}
    \caption{\textit{Left panel:} PDF of the square of the sound speed
      shown as a function of the energy density. Red-solid
      (orange-dashed) lines mark the $90\%$-confidence levels of the PDF,
      while the solid and dashed black (purple) lines are the $90\%$
      confidence levels of the PDFs relative to the centers of NSs with
      $M=1.4~M_\odot$ (or maximally massive, with $M=M_{\rm TOV}$) for
      the VL and the CL approaches, respectively. \textit{Right panel:} the same 
      as in the left panel but for the PDF of the pressure as a function of the
      energy density. In addition, marked by the red-dotted (orange-dotted)
      line is the $100\%$-confidence level for the VL (CL) method, while
      the uncertainties in CET~\citep{Hebeler2013} and
      pQCD~\citep{Fraga2014} are shown as blue and green areas,
      respectively.}
    \label{fig:EOS}
\end{figure*}

\section{Results}
\label{sec:result}

We next present the results of our Bayesian analyses focusing our
attention on the comparison between the VL and CL approaches. In
particular, we will concentrate on the EOS properties in
Sec.~\ref{sec:EOS} and on the NS properties in Sec.~\ref{sec:NS}. The
comparison will be based on contrasting the $90\%$-confidence intervals
and the medians of the probability density functions (PDFs) obtained from
the two analyses. For each quantity, we approximate the PDFs by counting
the number of curves that cross each cell of equally spaced grids and
then normalize them by the maximum count on the whole grid. Also, for
readability of the figures, we only display with colormaps the PDFs for
the VL analysis, referring the reader to~\citet{Altiparmak:2022} and
to~\citet{Ecker:2022xxj} for the corresponding PDFs in the case of the CL
analysis.

\subsection{EOS Properties}
\label{sec:EOS}

In this section we first analyze the EOS properties that can be inferred
from our two Bayesian setups. We first show in the left panel of
Fig.~\ref{fig:EOS} the PDFs of the sound speed that we computed using the
VL approach. The PDFs show a structure very similar to the ones computed
by~\citet{Altiparmak:2022} and \citet{Marczenko:2022}, that were obtained
making use of the CL approach. In particular, it is possible to note that
the sound speed rises rapidly till energy densities $e \approx 600~{\rm
  MeV}/{\rm fm}^{3}$, where it significantly exceeds the conformal limit
$c_{\rm CFT}^2=1/3$. This is a well-known feature that is necessary to
explain maximum-mass measurements $M\approx 2~M_\odot$~\citep[see
  discussion by][]{Bedaque2015, Hoyos:2016cob, Moustakidis2017,
  Kanakis-Pegios:2020jnf, Gorda:2022jvk, Brandes:2022nxa} that require a
large sound speed at moderate densities, and is further enhanced if
larger maximum-mass constraints are imposed, namely if $M_{\rm TOV}>
2~M_\odot$~\citep[see][]{Ecker:2022dlg}.

The decrease of the sound speed after the maximum at higher densities is
a consequence of the perturbative-QCD constraint imposed at very large
energy densities, as well as of the tidal-deformability constraint from
GW170817, that disfavors stiff models with too large sound speed.
Overall, when comparing the $90\%$-confidence level contours for the VL
and CL approaches (red-solid and orange-dashed lines, respectively), it
is apparent that not only the qualitative behavior is remarkably
similar, but also the quantitative ones. These differences become more
pronounced when comparing the $100\%$-confidence level contours
(red-dotted and orange-dotted lines, respectively) as it is in the tails
of the likelihood functions that the largest differences are imprinted in
the two methods. More importantly, the statistical relevance of these
stellar models is commensurate to the very low probability with which
they appear.

We should note that there are also two artificial features in the PDF
that are due to the parameterization method. One is the local minimum in
the upper bound of the $90\%$-confidence level at $c^2_s \approx 0.7$ and
$e\approx 350~{\rm MeV}/{\rm fm}^{3}$, which is an artifact of fixing the
chemical potentials at constant values and which vanishes when sampling
the chemical potentials randomly. The second one is the pronounced
maximum at large densities, close to where the perturbative QCD boundary
conditions are imposed. This maximum is caused by cases in which the
pressure at large densities is relatively low, but where it is still
possible to reach the allowed interval for $p_{\rm QCD}$ in a causal way
with strongly varying sound speed slightly below $\mu=2.6\, \rm
GeV$~\citep[see also][where this is discussed in more
  detail]{Altiparmak:2022}.

\begin{table*}
\begin{ruledtabular}
  \centering
  \caption{Neutron-star and EOS properties ($90\%$-confidence levels)
    from the variable likelihood (VL) and the constant likelihood (CL)
    methods. Listed are the radii $R$ of NSs with mass $M=1.4,
    2.0\,M_{\odot}$, with maximum mass $M=M_{\rm TOV}$, the corresponding
    tidal deformabilities $\Lambda_{1.4}$ and $\Lambda_{\rm TOV}$, as
    well as the number density $n_{\rm c}$, the pressure $p_{\rm c}$ and
    the sound speed $c^2_{s,c}$ at the center of typical
    ($M=1.4\,M_{\odot}$) and maximally massive ($M=M_{\rm TOV}$) NSs.}
\label{tab:compare}
\renewcommand{\arraystretch}{1.3}
\begin{tabular}{cccccccccccc}
	Method  &  $R_{\rm 1.4}$ &  $R_{\rm 2.0}$  &  $R_{\rm TOV}$ & $\Lambda_{\rm 1.4}$  &  $\Lambda_{\rm TOV}$  &  ${n_{\rm c, 1.4}}$   &  ${n_{\rm c, TOV}}$ &  $p_{\rm c, 1.4}$ &  $p_{\rm c, TOV}$ &  $(c^2_{s,c})_{1.4}$  &  $(c^2_{s,c})_{\rm TOV}$     \\
	&  $[{\rm km}]$ &  $[{\rm km}]$  &  $[{\rm km}]$  &  &  & $[{n_s}]$ & $[{n_s}]$ &  $[{\rm MeV}/{\rm fm}^3]$ &  $[{\rm MeV}/{\rm fm}^3]$ &  &     \\
        \hline
        VL      &  $12.2_{-0.9}^{+0.9}$      &  $12.4_{-1.1}^{+1.0}$       &  $11.8_{-1.0}^{+1.3}$       &  $450_{-190}^{+230}$ &  $11_{-5}^{+18}$  &  $2.4_{-0.5}^{+0.7}$             &  $5.5_{-1.2}^{+1.2}$            &  $60_{-20}^{+30}$                                &  $350_{-170}^{+210}$                             &  $0.60_{-0.22}^{+0.23}$ &  $0.23_{-0.17}^{+0.38}$ \\
        CL      &  $12.4_{-1.0}^{+0.5}$      &  $12.6_{-1.4}^{+0.7}$       &  $12.1_{-1.3}^{+1.1}$       &  $490_{-220}^{+130}$ &  $13_{-7}^{+26}$ & $2.3_{-0.4}^{+0.8}$             &  $5.2_{-1.1}^{+1.5}$            &  $50_{-10}^{+30}$                                &  $310_{-150}^{+220}$                             &  $0.63_{-0.25}^{+0.23}$ &  $0.20_{-0.16}^{+0.38}$ \\
\end{tabular}
\end{ruledtabular}
\end{table*}

Also shown in the left panel of Fig.~\ref{fig:EOS}, respectively with
solid and dashed black (purple) lines are the $90\%$ confidence levels of
the PDFs relative to the centers of NSs with $M=1.4~M_\odot$ (or
maximally massive, with $M=M_{\rm TOV}$) for the VL and the CL
approaches, respectively. Also in this case, it is possible to note that the
differences between the two evaluations of the likelihood functions are
very small. More specifically, we find that in the VL (CL) approach the
median value of the sound speed at the center of the NSs is
$(c_{s,\,c})^2_{1.4}=0.60~(0.63)$ for the typical mass $M=1.4\,M_{\odot}$
and that $97\%$ of the EOSs have central sound speeds
$c_{s,\,c}^2>1/3$. Similarly, we find that at the center of maximally
massive NSs with $M=M_{\rm TOV}$, the sound speed are
$(c_{s,\,c})^2_{\rm TOV}=0.23~(0.20)$, in agreement with the results
of~\citet{Ecker:2022dlg}, while only $28~(24)\%$ of the EOSs have central
sound speeds $c_{s,\,c}^2>1/3$; a detailed description of the results in
Fig.~\ref{fig:EOS} can be found in Table~\ref{tab:compare}.

In full similarity, we show in the right panel of Fig.~\ref{fig:EOS} the
two PDFs relative to the EOSs. Also in this case, the features of the PDF
of the VL approach is very similar to that obtained previously via the
CL method~\citep{Altiparmak:2022}. As remarked in several previous
studies, a most interesting feature is the clear change of slope (also
referred to as a ``kink'') of the PDF at energy densities $e \approx
600~{\rm MeV}/{\rm fm}^{3}$.~\citet{Annala2019} have interpreted this
feature as a sign for the appearance of deconfinement phase transition
from dense baryonic matter to quark matter (see also~\cite{Motta:2020xsg}
for an alternative interpretation in terms of hyperons). The
$90\%$-confidence intervals of the two approaches are again shown by
red-solid and orange-dashed lines and they agree remarkably well. Again,
deviations can be appreciated when comparing the $100\%$-confidence
intervals, marked by red and orange dotted lines, which exhibit
differences at energy densities $e \approx 300~{\rm MeV}/{\rm fm}^{3}$.
The origin of these differences is reasonably well understood. In
particular, the difference in the upper bound of these contours is due to
the diverse treatment of the GW constraint: the VL approach allows for
larger variations than the CL method since in the latter the binary tidal
deformability is assumed to be strictly $\tilde \Lambda<720$. As a
result, the VL approach allows for EOSs that can have radii for stars
$M\approx 1.4~M_\odot$ that are larger than those in the CL method (this
point will be discussed also in the next section).

Finally, the right panel of Fig.~\ref{fig:EOS} also reported by black
(purple) solid and dashed lines the values of the pressure and energy
density reached at the center of typical (maximally massive) NSs with
$M=1.4~M_\odot$ ($M_{\rm TOV}$) for the VL and the CL method,
respectively. Also in this case, the two sets of curves are almost
identical. These contours are useful to determine the median values of
the energy density and pressure at the center of representative
stars. More specifically, within the VL (CL) approach we find $e_{\rm
  c}=390~(380)~\rm MeV/fm^{3}$ and $p_{\rm c}=60~(50)~\rm MeV/fm^{3}$ for
typical mass ($M=1.4\,M_{\odot}$) NSs, while $e_{\rm c}=1100~(1030)~\rm
MeV/fm^{3}$ and $p_{\rm c}=350~(310)~\rm MeV/fm^{3}$ inside maximally
massive NSs ($M=M_{\rm TOV}$) (see also Table~\ref{tab:compare}).

\begin{figure*}[htb]
  \center
  \includegraphics[width=0.48\textwidth]{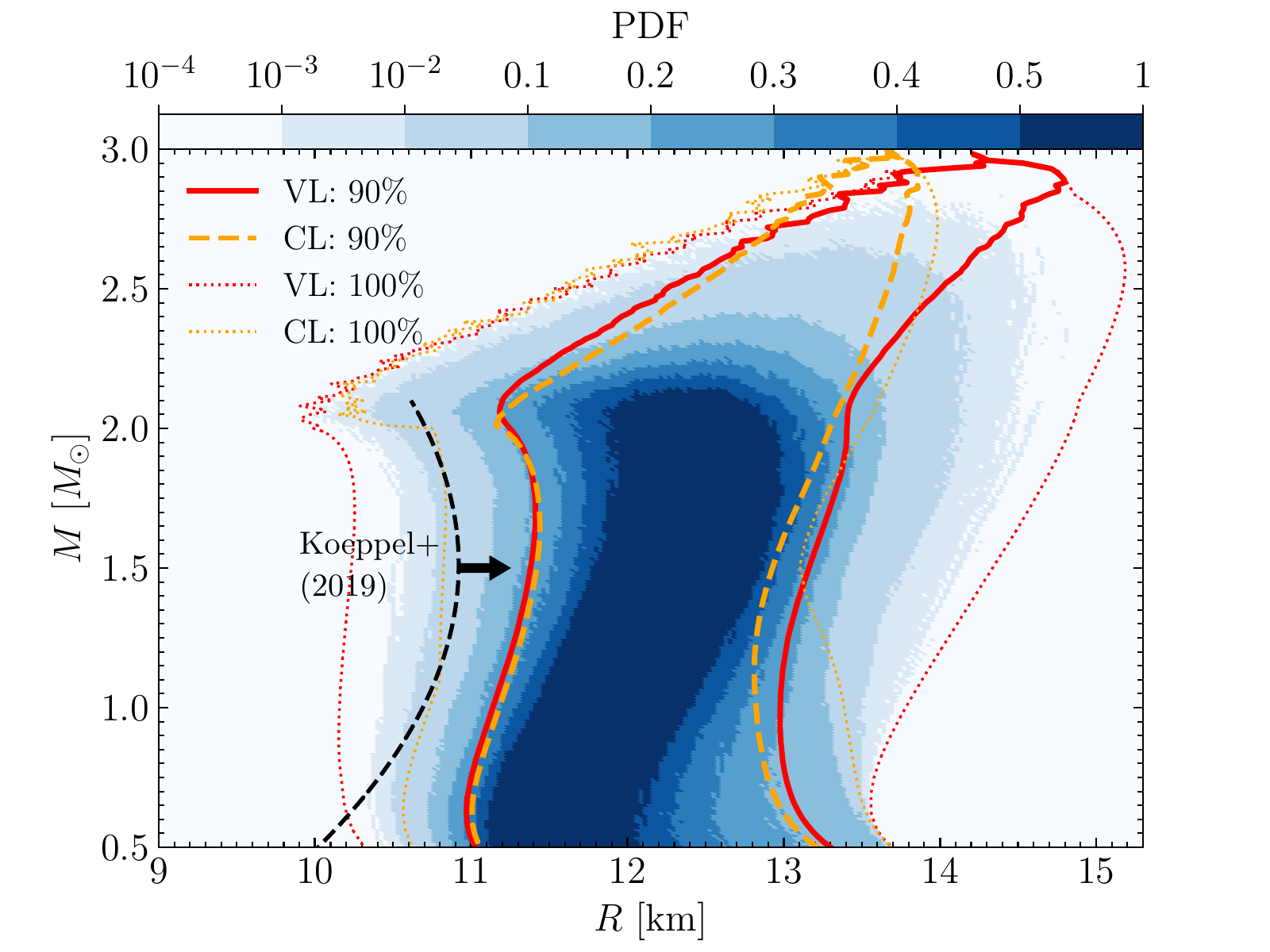}
  \includegraphics[width=0.48\textwidth]{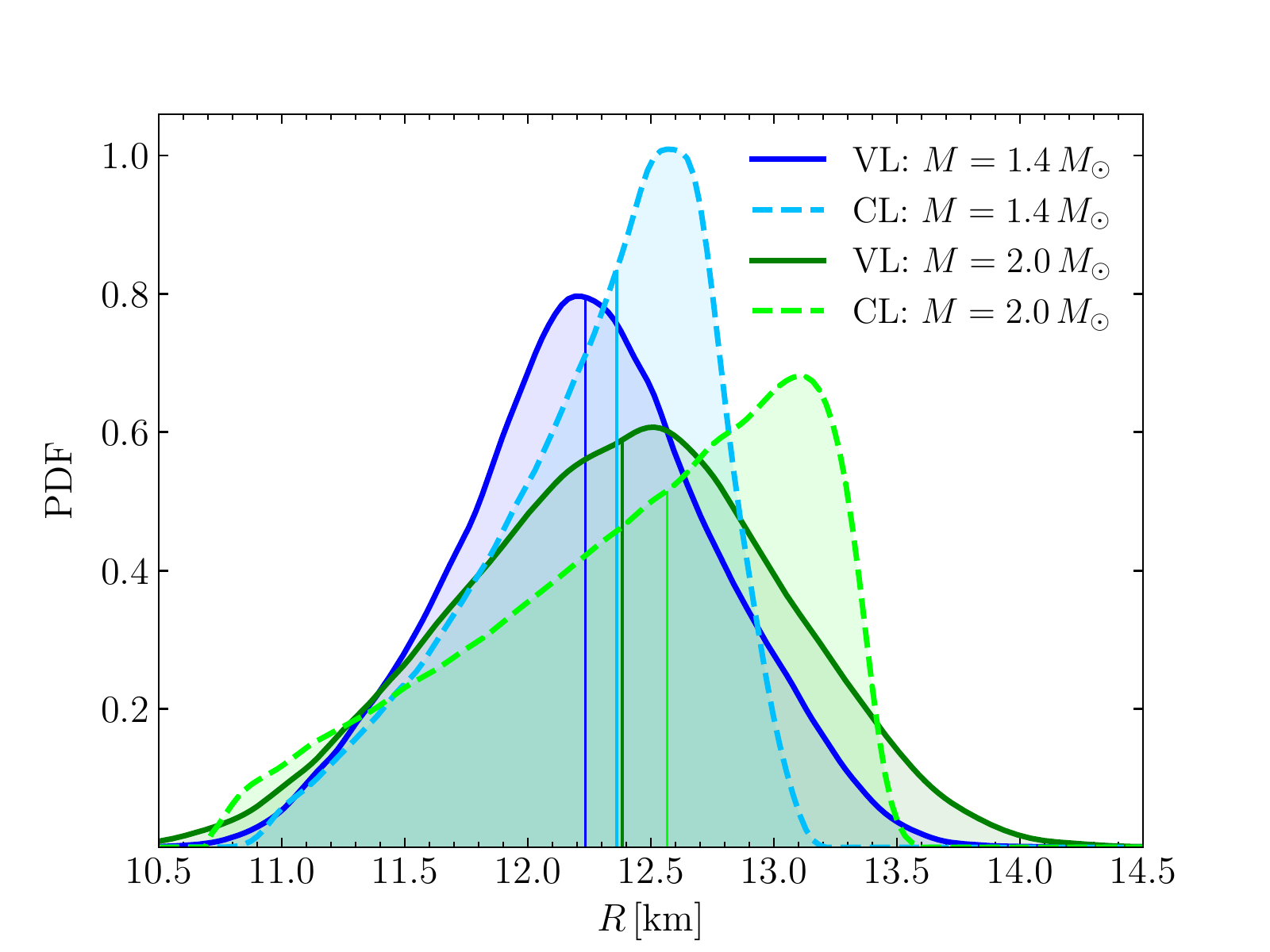}
  \caption{\textit{Left panel:} the same as in the right panel of
    Fig.~\ref{fig:EOS} but for the PDF of the stellar radius shown as a
    function of the stellar mass. In addition, marked by a black-dashed
    line is the lower bound for the radii as computed from considerations
    on the threshold mass~\citep{Koeppel2019}. \textit{Right panel:}
    One-dimensional PDF slices at fixed NS mass. The blue and green solid
    (light blue and light-green dashed) lines report the PDF cuts for the
    radii of a NS with $M=1.4\,M_{\odot}$ and $M=2.0\,M_{\odot}$ for the
    VL (CL) analysis, respectively. The vertical thin solid lines mark
    the corresponding median radii for the four PDFs.}
    \label{fig:MR}
\end{figure*}

\subsection{Neutron-Star Properties}
\label{sec:NS}

We next move on to the NS properties obtained from the VL and the CL
approaches. As in the previous section we show for clarity only the PDFs
for VL, but the confidence intervals for both VL and CL. Let us start
with a discussion of the mass-radius relation shown in
Fig.~\ref{fig:MR}. Overall, the PDFs from the VL and CL analyses are very
similar and centered around $R\approx 12\,{\rm km}$ for NSs with masses
$M\lesssim 2.0\,M_{\odot}$, while they start to differ for more massive
stars, where the VL approach allows for NSs with $R\sim 14\,{\rm km}$
although with rather small probability. The slightly wider PDF in the VL
approach can be explained by the fact that the NICER and the GW
constraints are imposed with distributions that have some support also
beyond the sharp-cutoff values that are employed in the CL approach. This
can be seen more clearly by comparing the outer contours ($100\%$
intervals) shown as red (VL) and orange (CL) dotted lines. As a result,
the VL PDF yields upper and lower bounds for the NS radii that are less
restrictive than those of the CL approach. On the other hand, when
comparing the $90\%$-confidence intervals of the two approaches (solid
and dashed lines) it is possible to conclude that they are essentially
identical, thus underlying that the methodological differences in the use
of the likelihood functions two approaches do not result in significant
changes in the PDFs. Furthermore, both $90\%$-confidence levels are in
good agreement with the analytic lower bound $R/{\rm km}\gtrsim 8.91
+2.66\,(M/M_{\odot})-0.88\,(M/M_{\odot})^2 $ (black-dashed line) derived
when using the detection of GW170817 and the estimates on the threshold
mass to prompt collapse~\citep{Koeppel2019, Tootle2021}.

Also shown in the right panel of Fig.~\ref{fig:MR} are the
one-dimensional PDF slices obtained in the two approaches for NSs with
fixed masses $M=1.4\,M_\odot$ and $M=2.0\,M_\odot$, together with the
corresponding median estimates (vertical solid lines) for their
radii. These cuts show more explicitly that the VL approach results in
distributions that are wider and less skewed than those for the CL case,
but also that the resulting median values differ only by $\approx
200~{\rm m}$ (see Table~\ref{tab:compare} for details); overall, the
differences between the two approaches are clearly much smaller than the
current observational uncertainties.

\begin{figure*}[htb]
    \center
    \includegraphics[width=0.50\textwidth]{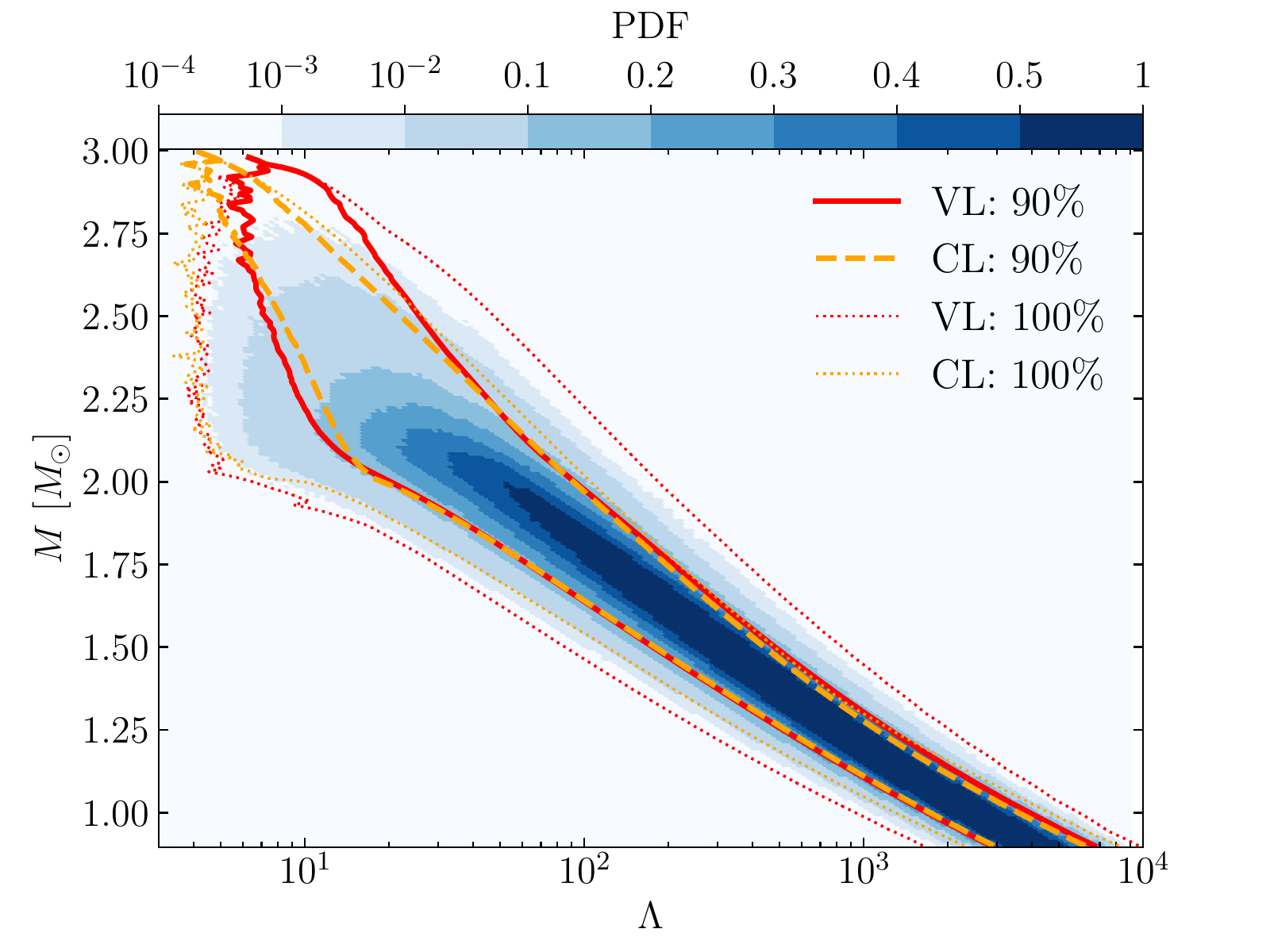}
    \includegraphics[width=0.46\textwidth]{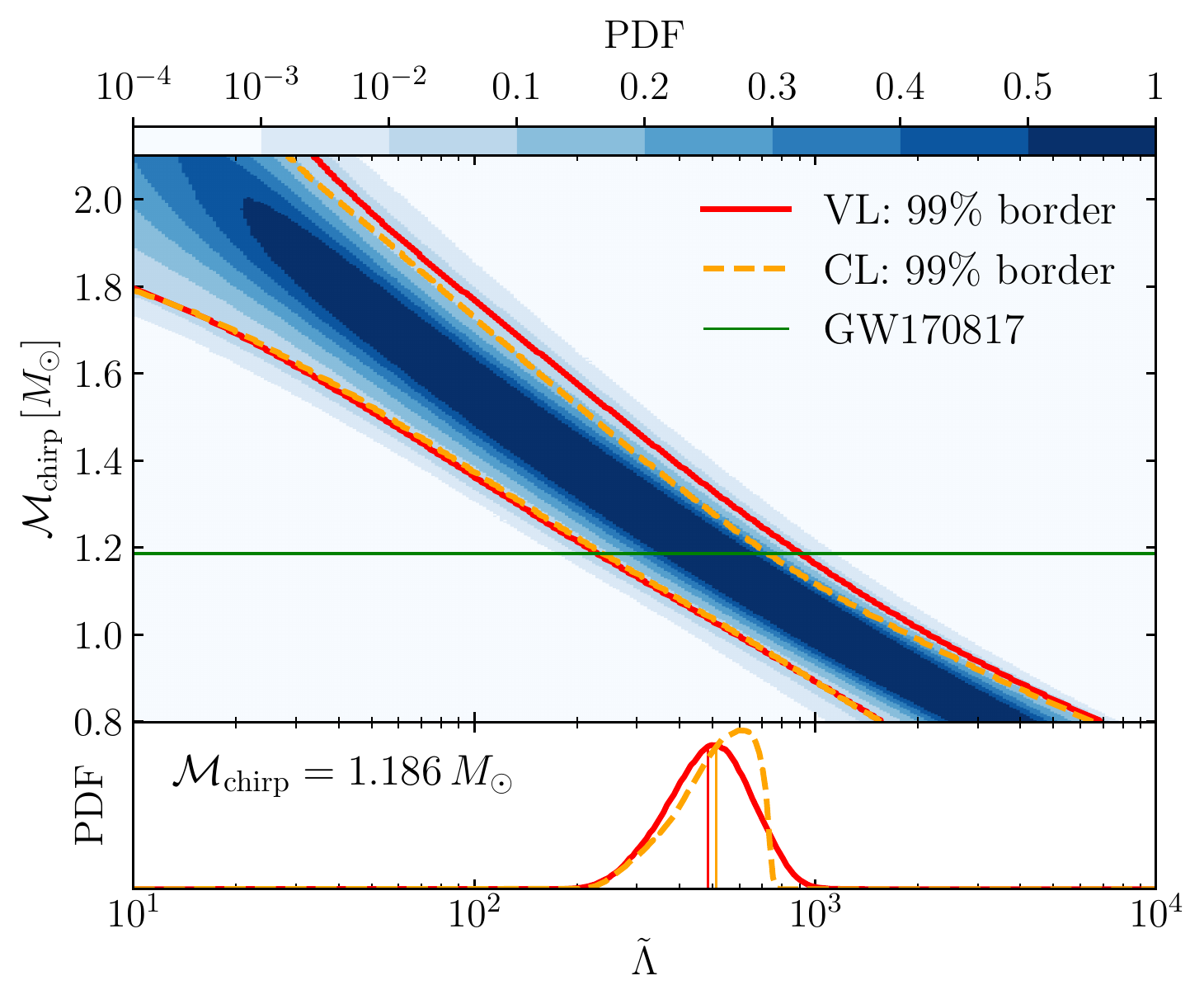}
    \caption{\textit{Left panel:} PDF of the tidal deformability
      $\Lambda$ of isolated NSs shown as a function of the mass (the
      color coding is the same as that in the previous
      figures). \textit{Right panel:} PDF of the binary tidal
      deformability $\tilde\Lambda$ of binary NS systems as a function of
      their chirp mass $\mathcal{M}_{\rm chirp}$. Indicated with a
      horizontal green line is the chirp mass of the GW170817 event,
      $\mathcal{M}_{\rm chirp}=1.186M_{\odot}$. The bottom part of the
      panel shows the corresponding one-dimensional PDF slices at
      $\mathcal{M}_{\rm chirp}=1.186M_{\odot}$, where the median values
      are marked by vertical thin lines. }
    \label{fig:MCL}
\end{figure*}

We next discuss the PDFs for the tidal deformabilities of isolated and
binary NSs that are plotted in Fig.~\ref{fig:MCL}. More specifically, in
the left panel we show how the PDFs of the tidal deformability $\Lambda$
of isolated NSs depends on the gravitational mass $M$ in these two
Bayesian analysis. There is a simple explanation for the overall trend of
the PDF: for generic EOSs that are viable, the relative change in the NS
radius is $\lesssim 6\%$ in the relevant mass range $M\approx
1-2\,M_\odot$, so that the tidal deformability is [see
  Eq.~\eqref{eq:def_lambda}] $\Lambda \sim k_2{M}^{-5} \sim {M}^{-p}$,
with $p\gtrsim 5$. Comparing the confidence intervals of the VL and CL
approaches leads to conclusions that are very similar to those drawn for
the mass-radius relation in Fig.~\ref{fig:MR}. More specifically, while
the $100\%$ confidence interval of the VL approach is significantly wider
than that of the CL method, the $90$\% contours are again almost
identical, so that the differences between the two methods affects mostly
the less probable and therefore less important regions of the PDFs.

In the right panel of Fig.~\ref{fig:MCL} we show instead the dependence
of the binary tidal deformability parameter $\tilde\Lambda$ on the chirp
mass $\mathcal{M}_{\rm chirp}$. The PDF is based on more than
$9.0\times10^{6}$ binary models with uniformly distributed mass ratios
$q\in[0.4, 1.0]$, i.e., a range that is reasonable according the mass
ratios in binary NS systems detected so far. Analogous results from the
CL approach have been shown by \citet{Altiparmak:2022} and were later
generalized to large-mass constraints by~\citet{Ecker:2022dlg}~\citep[see
  also][for similar studies but where no PDF is computed]{Zhao2018}. Note
that, in contrast to previous plots, the red-solid and orange-dashed
lines show here the $99\%$-confidence levels of the PDFs. As discussed by
\citep{Ecker:2022xxj}, these bounds -- and in the particular the lower
one -- are of great importance to constrain the tidal deformability (and
hence the EOS) using accurate measurements of the chirp mass of future GW
detections of binary NS mergers\footnote{We note that because
$\Lambda(M)$ and $\tilde{\Lambda}(\mathcal{M}_{\rm chirp})$ are related
(in the equal-mass case $\tilde{\Lambda}=\Lambda$ and $M_{\rm
  chirp}=M/\sqrt[5]{2}$), the left and right plots in Fig.~\ref{fig:MCL}
can be approximately mapped into each other after a simple rescaling
$M\to M/\sqrt[5]{2}$. This mapping is quite accurate for $q\approx 1$ but
degrades as the mass ratio is further decreased.}. Interestingly, the
lower bounds of these intervals are again almost identical, but the upper
bounds differ. The reason for this is that the upper bound is directly
set by the GW constraint of GW170817, while the lower bound mostly
depends on the TOV constraint~\citep{Ecker:2022dlg}.

\begin{figure*}[htb]
    \center
    \includegraphics[width=0.48\textwidth]{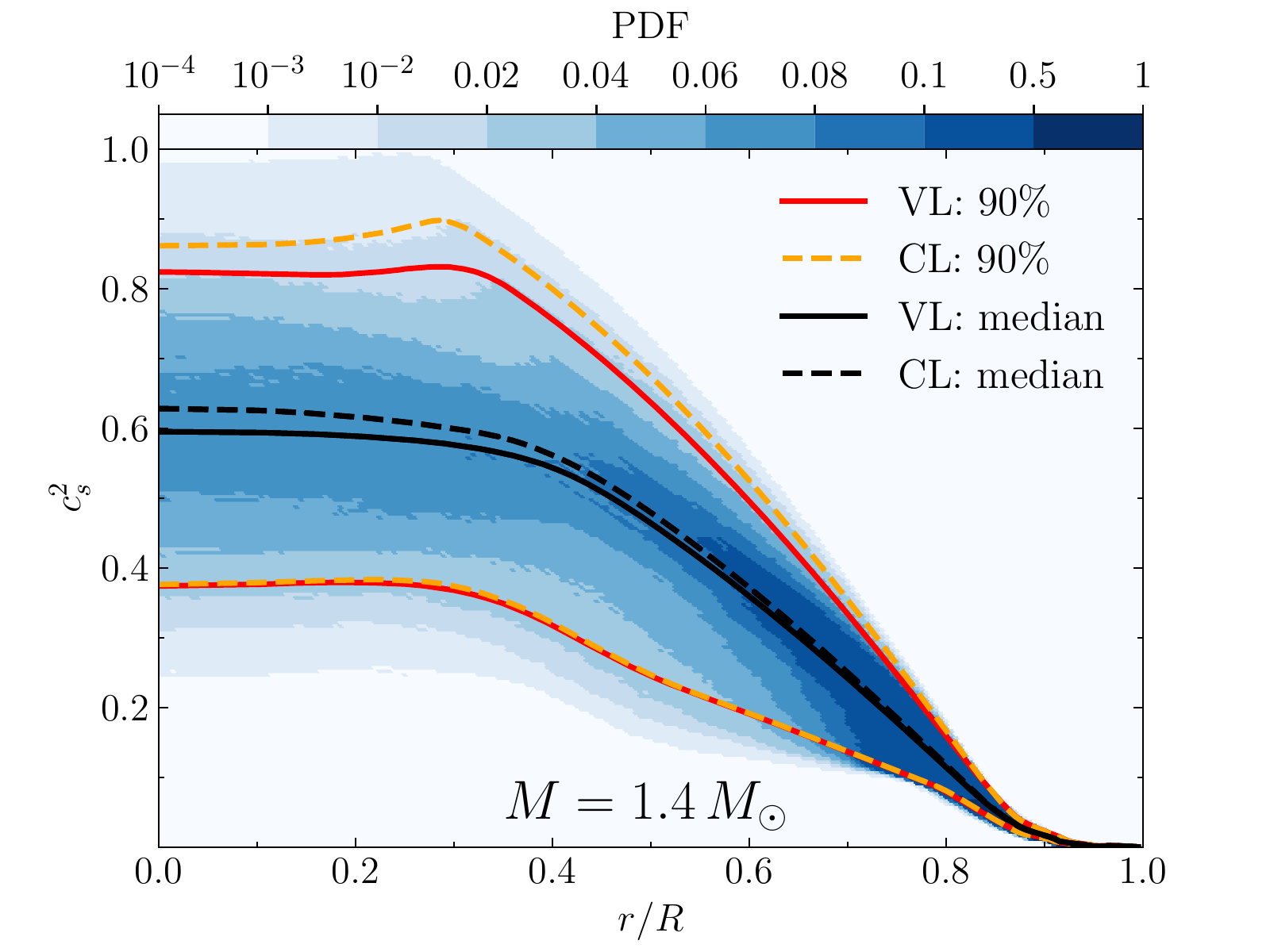}
    \includegraphics[width=0.48\textwidth]{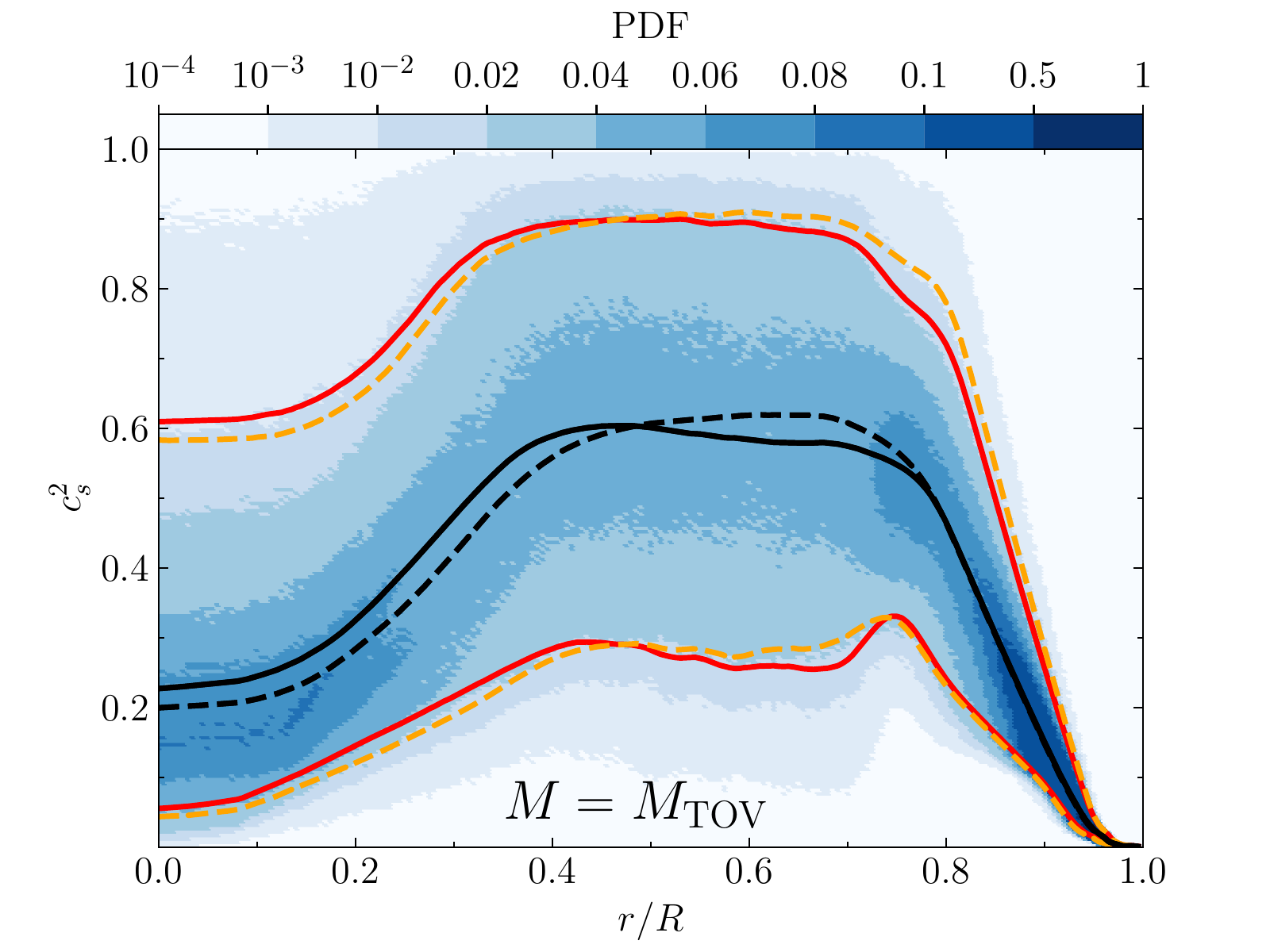}
    \caption{PDFs of the square of the sound speed as a function of
      normalized radial coordinate $r/R$ for NSs of mass
      $M=1.4\,M_{\odot}$ (left panel) and $M=M_{\rm TOV}$ (right
      panel). The color coding is the same as that in the previous
      figures, but the black-solid (black-dashed) report here the median
      for the VL (CL) approach.}
    \label{fig:radius_cs}
\end{figure*}

In particular, the sharp cutoff on $\tilde\Lambda$ at $\mathcal{M}_{\rm
  chirp}=1.186\,M_\odot$ employed in the CL approach strongly skews the
distribution towards the larger values of $\tilde\Lambda$. This is clearly 
visible in the bottom part of the right panel in Fig.~\ref{fig:MCL}, where we 
report cuts of the binary tidal deformability at $\mathcal{M}_{\rm
  chirp}=1.186M_{\odot}$ for the two approaches; clearly, the CL cut has
a rapid fall-off for $\tilde\Lambda \simeq 750$, while the VL cut is
almost Gaussian with a longer tail up to $\tilde\Lambda \simeq 1000$. As
a result, the CL approach slightly underestimates the upper bound on
$\tilde\Lambda$, while the VL approach gives a more conservative
estimate. Following~\citet{Altiparmak:2022}, we have calculated an
analytic fitting function that approximate these bounds and is given
by\footnote{For clarity, we do not report the fitting functions in
Fig.~\ref{fig:MCL}, but the quality of the fit can be appreciated from
Fig.~4 of~\citet{Altiparmak:2022}.}
\begin{equation}
  \label{eq:mclt}
  \tilde{\Lambda}_{\rm min(max)} = a+b\mathcal{M}_{\rm chirp}^c\,,
\end{equation}
where $a=-16,~b=560$ and $c=-5.1$ for the lower bound in both VL and CL
analyses, while $a=-19~(0.4), ~b=2200 ~(1900)$, and $c=-5.1~(-5.5)$ for the
upper bound in the VL~(CL) analysis. 

Notwithstanding these small differences, the median values (thin vertical lines)
for the case of the GW170817 event, whose chirp mass was
$\mathcal{M}_{\rm chirp}=1.186M_{\odot}$, are quite similar:
$\tilde\Lambda_{1.186}= 480^{+260}_{-190}$ (VL) versus
$510^{+180}_{-210}$ (CL). Furthermore, the
$\tilde{\Lambda}(\mathcal{M}_{\rm chirp})$ relation can also be used the
other way around. When in the future the EOS will be better constrained,
it will be then possible to constrain the source-frame chirp mass. Our
results can be combined with the detector frame chirp mass from the
waveform matched-filter to find the redshift of the luminosity distance
and thus constrain cosmological models~\citep[see, \eg][]{Messenger2013}.

We next switch our attention to the PDFs of the sound speed
$c_{s}^{2}(r/R)$ in the two approaches. This is presented in the left and
right panels of Fig.~\ref{fig:radius_cs}, where we report the spatial
dependence (\ie as a function of the normalized radial coordinate $r/R$)
of the PDFs of the sound speed inside typical ($M=1.4\,M_{\odot}$) and
maximally massive ($M=M_{\rm TOV}$) NSs.

The overall behavior of the PDF for the VL approach shown here is
similar to the one from the CL method presented by~\citet{Ecker:2022dlg,
  Ecker:2022xxj}. In particular, it is interesting to note that in
typical NSs the sound speed rises monotonically from the surface
towards the center, where it reaches values $(c^2_{s,\,c})_{1.4} \approx
0.6$ that are significantly larger than the conformal limit. On the other
hand, in maximally massive NSs the sound speed is non-monotonic and has
a local maximum with $c_{s,\rm max}^2>1/3$ in the outer layers (\ie $r/R
\sim 0.6$) and a local minimum at the center with $(c^2_{s,\,c})_{\rm
  TOV}\lesssim 1/3$ (see Table~\ref{tab:compare} for details). The
confidence intervals and median values from VL and CL approaches are
remarkably similar, even quantitatively. More specifically, for NSs with
mass $M=1.4~M_{\odot}$ (left panel) the VL prediction for the central
sound-speed, $(c^2_{s,\,c})_{1.4} = 0.60_{-0.22}^{+0.23}$, is slightly
lower than the one by CL, $(c^2_{s,\,c})_{1.4} =
0.63_{-0.25}^{+0.23}$. Note that also in this case, the major differences
are in the upper $90\%$-confidence levels, where the sharp cutoff on
$\tilde{\Lambda}$ employed in the CL analysis indirectly causes a slight
overestimate of the maximum possible values of the sound speed. Instead,
for maximally massive NSs with $M=M_{\rm TOV}$ (right panel), an opposite
behavior is observed: the central sound speed $(c^2_{s,\,c})_{\rm
  TOV}=0.23_{-0.17}^{+0.38}$, from the VL approach is slightly larger
than the one from the CL method, $(c^2_{s,\,c})_{\rm
  TOV}=0.20_{-0.16}^{+0.38}$. This is due to the pQCD constraint, which
introduces an anti-correlation between the sound-speed parameters at low
and high densities, namely: the large values of $c_{s}^{2}$ in the
low-density regions need to be compensated by lower values in the
high-density region.

\begin{figure}
    \center
    \includegraphics[width=0.48\textwidth]{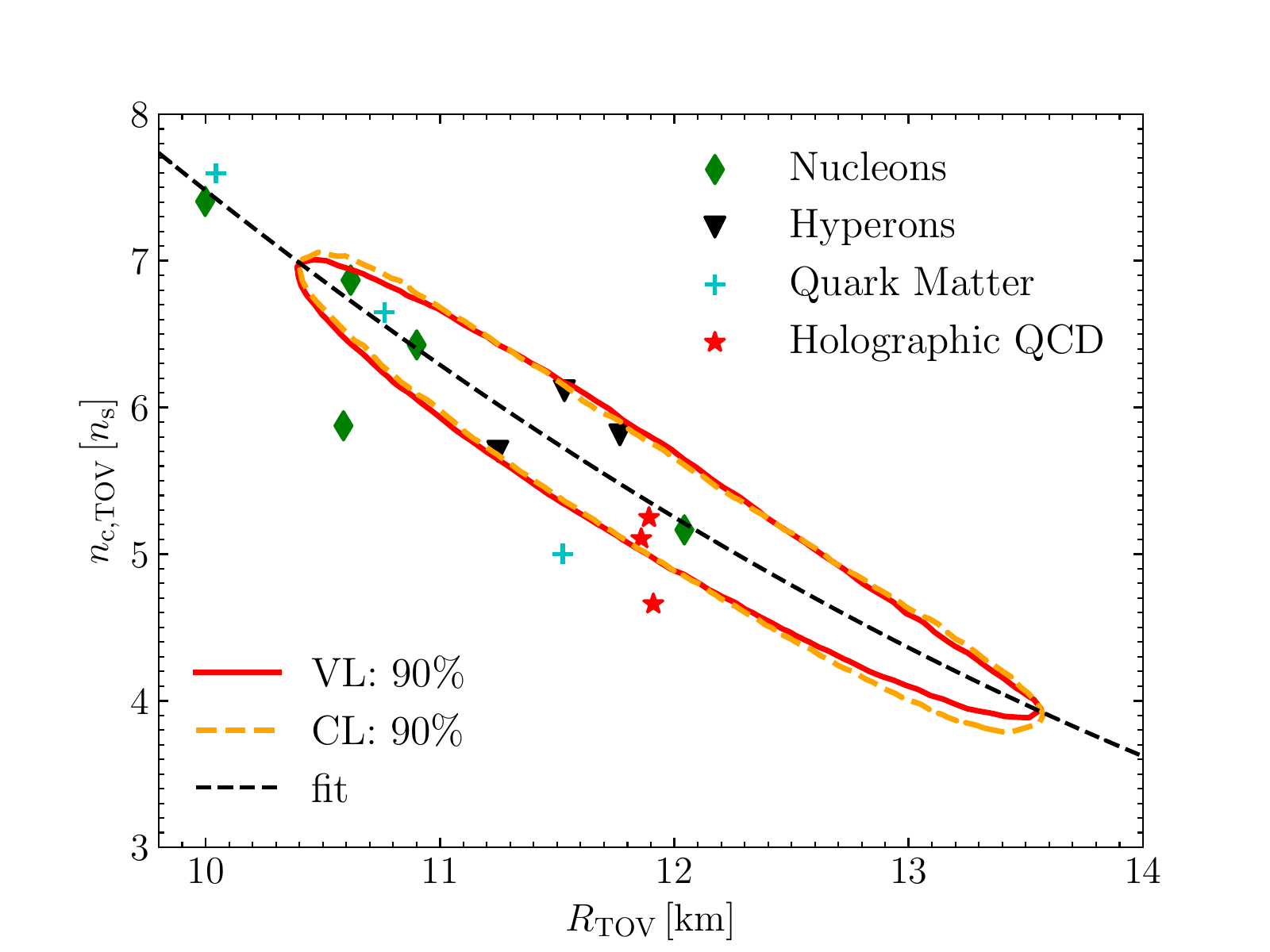}
    \caption{Relation between the normalized central number density of a
      maximally massive star, $n_{\rm c, TOV}/n_s$, and the corresponding
      radius, $R_{\rm TOV}$ (the color coding is the same as that in the
      previous figures). Also shown with various symbols are the
      corresponding values for EOSs of pure nuclear matter (green
      diamonds), with hyperons (black triangles), with quark matter (blue
      crosses) and models derived from holographic QCD (red stars). Shown
      instead with a black-dashed line is the quadratic fit given by
      Eq.~\eqref{eq:lower_lim_ctov}.} 
    \label{fig:uni_relation}
\end{figure}

Finally, we show in Fig.~\ref{fig:uni_relation} the correlation between
the radius $R_{\rm TOV}$ and the central number density $n_{\rm c, TOV}$
for an NS at the maximum mass $M_{\rm TOV}$. Here again, red-solid and
orange-dashed ellipses enclose the $90\%$-confidence levels obtained from
the VL and the CL approach, respectively. These intervals are essentially
identical, hence the relation between $R_{\rm TOV}$ and $n_{\rm c, TOV}$
is largely independent from the choice of the likelihood functions. In
addition, we show results for a number of micro-physical multi-purpose
EOS models of pure nuclear matter (``Nucleons''), with hyperons
(``Hyperons''), with quark matter (``Quark Matter'') and models derived
from holographic QCD. All EOSs are taken from the public database
CompOSE\footnote{CompOSE website:
\url{https://compose.obspm.fr/}.}~\citep{Typel2015} and correspond to
beta-equilibrium slices at the lowest available temperature with
electrons. Some of the corresponding properties are listed in
Table~\ref{tab:eos}.

As a concluding remark we note that the correlation between $R_{\rm TOV}$
and a normalized number density $n_{\rm c, TOV}/n_s$ is described by a
second-order polynomial
\begin{equation}
  \label{eq:lower_lim_ctov}
 \frac{n_{c,\,{\rm TOV}}}{n_s} = d_0~\left[1-\left(\frac{R_{\rm TOV}}{10\,
     \rm km}\right)\right] +d_1~\left(\frac{R_{\rm TOV}}{10\, \rm
   km}\right)^2\,,
\end{equation}
where the coefficients are given by $d_0=27.6$ and $d_1=7.5$. The
relative residuals are well described by a Gaussian distribution centered
on zero and have a standard deviation of $3.7\%$. The importance of
Eq.~\eqref{eq:lower_lim_ctov} is that it allows to estimate the maximum
central density from future radius measurements of very massive NSs. We
have found that a relation similar to Eq.~\eqref{eq:lower_lim_ctov} holds
also when considering the radii of NSs with $M=1.4\,M_{\odot}$. In this
case, the coefficients are given by $d_0=26.6$ and $d_1=7.6$, but the
scatter is four times larger, reaching deviations of $\simeq 30\%$;
although larger than in the case of maximally massive stars, these
deviations are much smaller than the uncertainty in ${n_{c,\,{\rm
      TOV}}}/{n_s} \simeq 3.7-7.0$.

\section{Conclusions}
\label{sec:conc_disc}

We have here investigated the systematic differences that are introduced
when performing a Bayesian-inference analysis of the EOS of neutron stars
employing either a variable (VL) or a constant (CL) likelihood. The
former is routinely adopted in Bayesian analyses as it allows to
introduce suitably variable likelihood functions for the set of
constraints in the model. The latter, on the other hand, mimics the
approach frequently used in those analyses where the data are used
as sharp cutoffs and uniform likelihoods. The advantages of the VL
method are therefore that it retains the full information on the
distributions of the measurements, thus making an exhaustive usage of the
data; its disadvantages are that it is more complex to implement and
computationally more expensive. The advantages of the CL method, on the
other hand, are to be found that the simplicity of its implementation and
the comparatively smaller computational costs; its disadvantages are
however that it does not fully exploit all the information from the
measurements and is not adequate when the measurement results exhibit a
bimodal or multi-modal behavior (a case not considered here).

In both approaches, the EOSs have been built using the sound-speed
parameterization method~\citep{Annala2019, Altiparmak:2022,
  Ecker:2022xxj}, with identical priors for the initial EOS ensemble. For
simplicity, we have restricted our comparison to observational
constraints only, meaning that the prior ensembles in both cases are
built by imposing the CET and perturbative QCD constraints in a
fixed-cutoff manner. As a result, the likelihood functions of the CET and
perturbative QCD constraints are constant, while that of the
observational constraints are suitably variable. Overall, the statistics
are performed making use of more than $1.5 \times 10^5$ causal and
thermodynamically consistent posterior EOSs that satisfy the nuclear
theory and perturbative QCD boundary conditions, as well as the
observational constraints from pulsars and GW measurements.

In our analysis we have considered how the two inferences differ when
considering either the properties of the EOS (\eg sound speed, pressure
behavior as a function of the energy density, and sound-speed variation
within the stellar interior) or of the NSs (masses, radii, and tidal
deformabilities). In all cases, we have found that the two approaches
lead to very similar results and that the $90\%$-confidence levels are
essentially overlapping. Some differences do appear when comparing the
$100\%$-confidence levels, but these differences concern regions of the
PDFs where the probability is extremely small, thus leading to the
conclusion that the role played by the functional form of the likelihood
function in our Bayesian inference is small.

When concentrating on what are the main sources of difference we note
that these can almost always be attributed to the sharp cutoff on
$\tilde{\Lambda}$ (\ie $\tilde \Lambda \leq 720$) employed in the CL
analysis, which indirectly causes a slight overestimate of the maximum
possible values of the sound speed in a range of energy densities inside
that allowed for plausible stellar models, but also a slight
underestimate of the maximum possible values of the stellar radii for all
masses. Finally, the sharp cutoff on ${\tilde \Lambda}$ also leads to a
smaller upper bound on $\tilde\Lambda$ with a consequent underestimate of
the binary tidal deformability using measured chirp mass.

In addition to the comparison between the two inference approaches and
the assessment of the role of the likelihood functions, our analysis has
also produced two noteworthy results. Firstly, a clear inverse
correlation between the normalized central number density of a maximally
massive star, $n_{\rm c, TOV}/n_s$, and the radius of either a typical
$M=1.4\,M_{\odot}$ NS or of the corresponding maximally massive star,
$R_{\rm TOV}$. Once a reliable measurement of one of these radii is
accomplished, the relation will provide a rather stringent estimate of
the number density that needs to be matched by studies building EOSs from
nuclear theory. Secondly, and most importantly, it has confirmed the
relation found between the chirp mass $\mathcal{M}_{\rm chirp}$ and the
binary tidal deformability $\tilde{\Lambda}$. We need to underline that
the importance of this result is that it relates a quantity that is
directly measurable -- and is very accurately measured -- from
gravitational-wave observations, $\mathcal{M}_{\rm chirp}$, with a
quantity that contains important information on the micro-physics
$\tilde{\Lambda}$. For example, when considering the case of the GW170817
event, whose chirp mass is $\mathcal{M}_{\rm chirp}=1.186M_{\odot}$, the
bounds obtained in the case of the VL approach are
$\tilde\Lambda_{1.186}= 480^{+260}_{-190}$.

\acknowledgments

We thank T. Gorda, Y.-Z. Fan, and M.-Z. Han for insightful discussions.
Partial funding comes from the State of Hesse within the Research Cluster
ELEMENTS (Project ID 500/10.006), by the ERC Advanced Grant ``JETSET:
Launching, propagation and emission of relativistic jets from binary
mergers and across mass scales'' (Grant No. 884631). CE acknowledges
support by the Deutsche Forschungsgemeinschaft (DFG, German Research
Foundation) through the CRC-TR 211 ``Strong-interaction matter under
extreme conditions''-- project number 315477589 -- TRR 211. JLJ
acknowledges support by the Alexander von Humboldt Foundation. LR
acknowledges the Walter Greiner Gesellschaft zur F\"orderung der
physikalischen Grundlagenforschung e.V. through the Carl W. Fueck
Laureatus Chair. The calculations were performed on the local ITP
Supercomputing Clusters Iboga and Calea.

\bibliographystyle{aasjournal}

\bibliography{thebibliography}

\newpage
\appendix

\section{Comparison to HESS~J1731-347}
\label{app:HESS}

As mentioned in the main text,~\citet{Doroshenko2022} have recently
reported the combined mass-radius measurement of a very light compact
object within the supernova remnant HESS~J1731-347 having a mass
$M=0.77^{+0.20}_{-0.17}~M_\odot$ and radius
$R=10.4^{+0.86}_{-0.78}~\,{\rm km}$. Because of the untypically small
values of mass and radius, that were deduced from combining distance
estimates from Gaia observations with X-ray spectrum modelling, the
nature and composition of this object is currently unclear. Indeed, the
compact star is speculated to be either the lightest (and smallest) NS
known or a ``strange star'' composed of more exotic constituents than
just nucleonic matter.

\begin{figure}
    \center
    \includegraphics[width=0.47\textwidth]{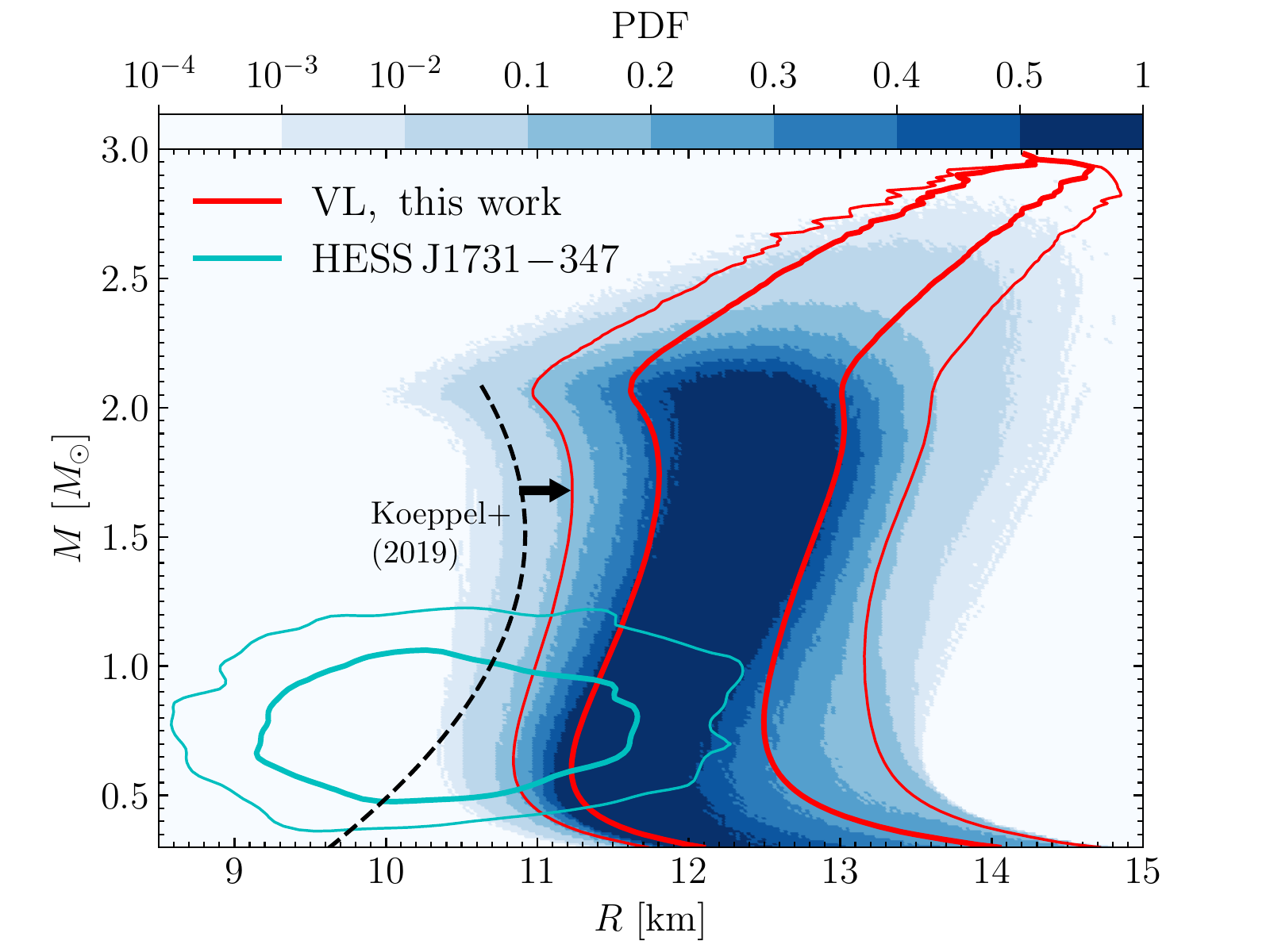}
    \includegraphics[width=0.49\textwidth]{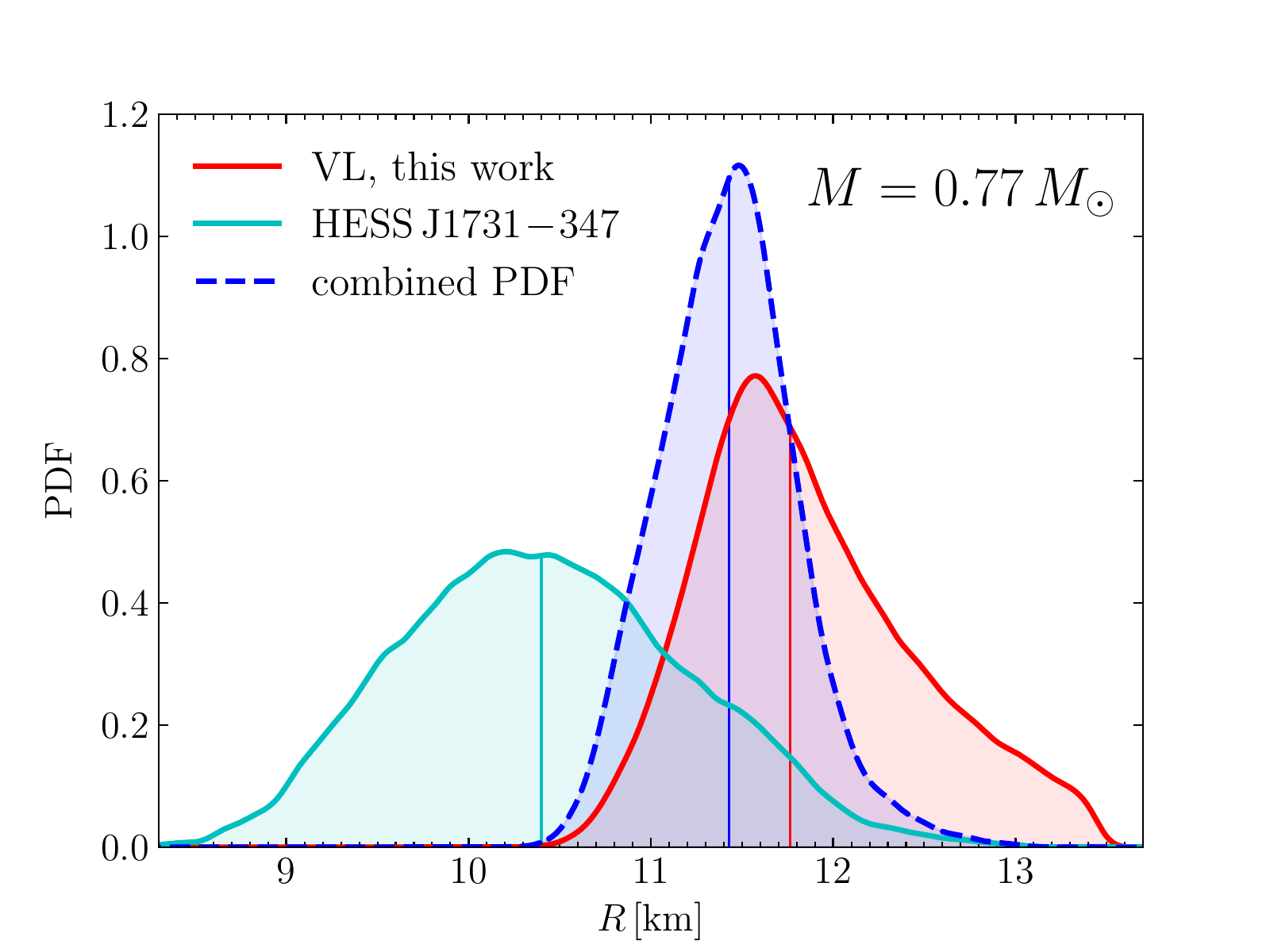}
    \caption{\textit{Left panel:} Comparison of the mass-radius data for
      HESS~J1731-347 and the mass-radius distribution of our VL analysis.
      Thin (thick) red lines represent the $95.4\%$ ($68.3\%$) confidence
      level, while light-blue contours mark the $95.4\%$ ($68.3\%$)
      measurement posterior samples of HESS~J1731-347. \textit{Right
        panel:} One-dimensional PDF slices at constant mass
      $M=0.77\,M_{\odot}$ using the same color convention as in the left
      panel. The blue-dashed line combines the two PDFs and provides our
      new estimate for the radius of HESS~J1731-347.}
    \label{fig:appendix}
\end{figure}

Assuming that HESS~J1731-347 is composed of standard nuclear matter and
maintaining the same uncertainties in the mass measurement as those
reported by~\citet{Doroshenko2022}, we can use the results of our
Bayesian analysis with VL to infer the distribution of stellar radii
compatible with our posterior PDFs. In particular, we show in the left
panel of Fig.~\ref{fig:appendix} the $68.3\%$ and $95.4\%$ confidence
levels of the PDFs for the mass and radius of NSs as obtained in our
analysis (thick and thin red solid lines, respectively) together with the
corresponding levels provided by~\citet{Doroshenko2022} (thick and thin
light-blue solid lines, respectively)\footnote{We use the
\href{https://zenodo.org/record/6702216}{posterior data}.} obtained from
the fitting of the X-ray data alone using a single-temperature carbon
atmosphere model and Gaia parallax priors. Already a crude comparison
reveals that our analysis tends to predict systematically larger radii
within the mass of interest, with a difference of about one kilometer. It
is a matter of concern that a good portion of the allowed space for the
mass and radius suggested by~\citet{Doroshenko2022} violates the
lower-limits on the radius set by threshold mass by
\citet{Koeppel2019}~\citep[see also the estimate of][which requires
  $R_{1.6}\geq 10.30\,{\rm km}$]{Bauswein2017b}, although it is fair to
recognise that such lower limits have been deduced for larger masses.

Our tendency to have larger radii can be seen more clearly in the right
panel of Fig.~\ref{fig:appendix}, where we compare the corresponding
one-dimensional PDF slices at $M=0.77~M_\odot$ from the analysis
of~\citet{Doroshenko2022} (light-blue solid line) and from our inference
(red solid line). Clearly, the corresponding median values can be seen to
differ by $\approx 1.5~\,{\rm km}$. Combining the two PDF slices and
properly normalizing leads to PDF shown with the blue dashed line, which
suggests a radius $R=11.43^{+0.64}_{-0.60}~\,{\rm km}$ in $90\%$ interval
and therefore about one kilometer larger than that deduced
by~\citet{Doroshenko2022}. It will be interesting to see if this
difference persists when the present (crude) comparison is further
refined by imposing in our Bayesian analysis the constraints coming from
the measurements of~\citet{Doroshenko2022} and by lower limits set
by~\citet{Koeppel2019}; we postpone this investigation to future work.

\section{EOSs employed in Figure~5}
\label{app:table}

For completeness, we provide here the basic properties of the EOSs used
as symbols in Fig.~\ref{fig:uni_relation}. These properties are listed in
Table~\ref{tab:eos} and we provide information on the NS radii ($R_{\rm
  1.4},~R_{\rm 2.0}$ at $M=1.4,~2.0~M_\odot$ and ${R_{\rm TOV}}$ at
$M=M_{\rm TOV}$), on the maximum mass $M_{\rm TOV}$ and on the binary
tidal deformability $\tilde \Lambda_{1.186}$ of GW170817-like binaries
with chirp mass $\mathcal{M}_c=1.186~M_\odot$ when considering two
different mass ratios, \ie $q=1,0.7$.

\begin{table*}[h]
\begin{ruledtabular}
  \centering
  \caption{Properties of the EOS models used in
    Fig.~\ref{fig:uni_relation}. The properties listed are the
    corresponding NS radii $R_{\rm 1.4},~R_{\rm 2.0}$ and ${R_{\rm TOV}}$
    at $M=1.4,~2.0~M_\odot$ and the maximum mass $M_{\rm TOV}$,
    respectively, as well as the binary tidal deformability parameter
    $\tilde \Lambda_{1.186}$ of GW170817 like binaries with chirp mass
    $\mathcal{M}_c=1.186~M_\odot$ for two different mass ratios
    $q=1,0.7$.}
\label{tab:eos}
\begin{tabular}{llccccccc}
	Type  & EOS    & ${R_{\rm 1.4}}$ & ${R_{\rm 2.0}}$ &  ${R_{\rm TOV}}$ &  ${M_{\rm TOV}}$  & $\tilde{\Lambda}^{\rm q=1~(0.7)}_{1.186}$  &   Ref. \\
             &            &  $[{\rm km}]$     &  $[{\rm km}]$      &  $[{\rm km}]$        &  $[{M_{\odot}}]$     &   &   &   \\
  \hline  
Nucleons   &  DD2                  &  $13.76$  &  $13.45$  &  $12.04$  &  $2.42$  &  $816 ~(784)$   &   \cite{Hempel2009}       \\
             &  LS220                   &  $12.66$  &  $11.34$  &  $10.62$  &  $2.04$  &  $574~(569)$  &   \cite{Schneider2017}        \\
             &  SFHx                 &  $12.4$   &  $11.78$  &  $10.9$   &  $2.13$  &  $464 ~(451)$   &   \cite{Steiner2013}      \\
             &  SLy4                 &  $11.73$  &  $10.67$  &  $10.0$   &  $2.05$  &  $313 ~(309)$   &   \cite{Schneider2017}    \\
             &  KOST2                &  $11.58$  &  $11.21$  &  $10.2$   &  $2.22$  &  $358 ~(352)$   &   \cite{Togashi2017}      \\
  \hline
Hyperons     &  DD2$\Lambda \phi$    &  $13.75$  &  $12.84$  &  $11.77$  &  $2.1$   &  $813 ~(772)$   &   \cite{Banik2014}        \\
             &  DD2Y                 &  $13.75$  &  $12.33$  &  $11.53$  &  $2.04$  &  $814 ~(767)$   &   \cite{MARQUES2017}      \\
             &  DD2Y$\Delta$ 1.1-1.1 &  $13.47$  &  $12.22$  &  $10.63$  &  $2.17$  &  $698 ~(687)$   &   \cite{Raduta2020}       \\
  \hline
Quark Matter &  RDF1.6               &  $12.46$  &  $10.31$  &  $10.04$  &  $2.01$  &  $509 ~(501)$   &   \cite{Bastian:2020unt}  \\
             &  RDF1.7               &  $12.46$  &  $11.44$  &  $10.76$  &  $2.11$  &  $508 ~(506)$   &   \cite{Bastian:2020unt}  \\
             &  TM1B145              &  $12.92$  &  $12.25$  &  $10.57$  &  $2.27$  &  $625 ~(612)$   &   \cite{Sagert2010}       \\
             &  V-QCD soft           &  $12.42$  &  $12.01$  &  $11.91$  &  $2.02$  &  $537 ~(517)$   &   \cite{Demircik:2021zll} \\
             &  V-QCD interm         &  $12.51$  &  $12.4$   &  $11.86$  &  $2.14$  &  $565 ~(543)$   &   \cite{Demircik:2021zll} \\
             &  V-QCD stiff          &  $12.65$  &  $12.85$  &  $11.89$  &  $2.34$  &  $617 ~(591)$   &   \cite{Demircik:2021zll} \\
\end{tabular}
\end{ruledtabular}
\end{table*}

\end{document}